\documentclass[usenatbib]{aa}  

\usepackage{graphicx}
\usepackage{subcaption}

\usepackage{txfonts}
\usepackage{xcolor}

\usepackage[colorlinks=true,linkcolor=blue,filecolor=magenta,urlcolor=cyan,citecolor=blue]{hyperref}

\providecommand{\teff}{$T_{\rm eff}$}
\providecommand{\feh}{[Fe/H]}
\providecommand{\logg}{$\log g$}

\begin{document}

   \title{Host-star and exoplanet composition:\\\
   Polluted white dwarf reveals depletion of moderately refractory elements in planetary material\thanks{This paper includes data gathered with the 6.5 meter Magellan Telescopes located at Las Campanas Observatory, Chile.}}
   \titlerunning{Polluted white dwarf reveals depletion of moderately refractory elements in planetary material}

   \author{Claudia Aguilera-Gómez \inst{1, 2} \and
          Laura K. Rogers \inst{3} \and 
          Amy Bonsor \inst{3} \and
          Paula Jofré \inst{4,2} \and
          Simon Blouin \inst{5} \and
          Oliver Shorttle \inst{3} \and
          Andrew M. Buchan \inst{6}\and
          Yuqi Li \inst{3} \and
          Siyi Xu \inst{7}}

   \institute{Instituto de Astrofísica, Pontificia Universidad Católica de Chile, Av. Vicuña Mackenna 4860, 782-0436 Macul, Santiago, Chile\\
              \email{craguile@uc.cl}
         \and
            Millennium Nucleus ERIS
        \and
            Institute of Astronomy, University of Cambridge, Madingley Road, Cambridge CB3 0HA, UK
        \and
            Instituto de Estudios Astrofísicos, Universidad Diego Portales, Av. Ejército Libertador 441, Santiago, Chile
        \and
            Department of Physics and Astronomy, University of Victoria, Victoria, BC V8W 2Y2, Canada
        \and
            Department of Physics, University of Warwick, Coventry CV4 7AL, UK
        \and
            Gemini Observatory/NSF’s NOIRLab, 670 N. A’ohoku Place, Hilo, HI 96720, USA
             }

   \date{Received X, 2024; accepted X, 2024}

 
  \abstract
   {Planets form from the same cloud of molecular gas and dust as their host stars. Confirming if planetary bodies acquire the same refractory element composition as their natal disc during formation, and how efficiently volatile elements are incorporated into growing planets, is key to linking the poorly constrained interior composition of rocky exoplanets to the observationally-constrained composition of their host star. Such comparisons also afford insight into the planet formation process.}
   {This work compares planetary composition with host-star composition using observations of a white dwarf that has accreted planetary material and its F-type star wide binary companion as a reference for the composition of the natal molecular gas and dust.}
   {Spectroscopic analysis reveals abundances of Fe, Mg, Si, Ca, and Ti in both stars. We use the white dwarf measurements to estimate the composition of the exoplanetary material and the F-type companion to constrain the composition of the material the planet formed from.}
   {Comparing planetary material to the composition of its natal cloud, our results reveal that the planetary material is depleted in moderate refractories (Mg, Si, Fe) relative to the refractory material (Ca, Ti). Grouping elements based on their condensation temperatures is key to linking stellar and planetary compositions.}
   {Fractionation during formation or subsequent planetary evolution leads to the depletion of moderate refractories from the planetary material accreted by the white dwarf. This signature, as seen for bulk Earth, will likely be present in the composition of many exoplanets relative to their host-stars.} 
   

   \maketitle
%

\section{Introduction}

The intricate process of planet formation remains one of the most challenging frontiers of modern astrophysics. Stars and planets form from the collapse of a cloud of interstellar gas and dust \citep{Wang2019composition, Adibekyan2021}. As planets form, volatile elements, characterized by low condensation temperatures, are not present in solid form at high temperatures close to the host star. In contrast, the most refractory elements remain in solid form and likely present in growing planets in roughly the same relative proportions as in the star. Moderately volatile and refractory elements fall between these extremes, and their presence in growing planets will be sensitive to the temperature of their formation environment. Additionally, subsequent processing of the planet during the post-nebular phase could further drive the loss of moderately volatile and moderately refractory elements due to impacts and melting.
These processes imply that the composition of an exoplanet, unknown in detail, cannot be simply linked to that of its host star. Exoplanet detection techniques yield a planet's mass and radius and can only provide information on the bulk density, leaving degeneracies in determining its true composition \citep[e.g.][]{seager2007mass,Dorn2015}. 
As the composition of a planet is paramount for unraveling its interior structure, there is a challenge in using stellar compositions to constrain planet structure. Better insights into how stellar compositions, or the composition of the primordial material, map to planetary compositions would alleviate this.

As host stars and their planets form from the same cloud of interstellar gas, the host stars are invaluable objects for understanding the composition of exoplanets. In particular, the chemical composition of a star provides crucial insights into the primordial material available to form the exoplanets \citep{Thiabaud2015}. A valuable approach, then, is comparing the host star's composition with that of the bulk planet composition in the context of our Solar System. Here, the analysis of meteorites provides a window into the study of rocky planetary material. The abundance of refractory elements, such as Ca, Al, and Ti, in chondritic meteorites, agrees with the composition of the Sun \citep{AndersEbihara1982}, although with slight variations in elemental ratios, underscoring the complexity of the planet formation process. In contrast, there is a depletion of volatile elements. \citet{WangLineweaverIreland2018} documents that moderately refractory elements like Si, Mg, Fe, and Ni exhibit subtle differences between the Sun and the Earth, while elements such as C and S experience substantial depletion.

White dwarfs (WDs) present and additional avenue for acquiring the bulk composition of planetesimals. These stellar remnants do not retain the chemical characteristics of their progenitors and typically exhibit pure hydrogen or helium atmospheres \citep[e.g.][]{koester2009}. Heavier elements, on the other hand, are expected to sink out from the visible layers in timescales much shorter than the cooling age of the WD.
As such, the presence of metals in the spectra of WDs serves as an indication of accretion from smaller rocky bodies that have endured the host star's evolution and subsequently been scattered inwards, undergoing tidal disruption by the WD \citep{Veras2016}. Approximately 20--50\% of WDs display signs of pollution \citep[e.g.][]{koester2014frequency}, with compelling evidence suggesting the ingestion of planetary material to account for this phenomenon \citep{Jura2003}. This hypothesis is further supported by the evidence for circumstellar discs of gas and dust \citep[e.g.][]{gaensicke2006gaseous,Farihi2016}, transits of disrupted bodies \citep{Vanderburg2015, Guidry2021}, and X-ray signals indicative of ongoing accretion \citep{Cunningham2022}.

These metal polluted WDs have served as valuable tools for studying the composition of accreted planetesimals and delving into the geology of exoplanetary material \citep[e.g.][]{Harrison2018,Swan2023}. The abundance of specific elements detected in the WD atmosphere offers crucial insights into the process of planet formation itself and of the subsequent heating of the planetesimal. Analyzing the amount of volatiles provides estimations of the planets' formation radius in the disc, the ratio of different types of elements aids in understanding the formation of planetary cores and crusts, and the accretion of icy objects provides insights into the cores of giant planets \citep[e.g.][]{Harrison2018, gansicke2019accretion,harrison2021bayesian}.

Host star compositions are often used as a proxy for exoplanetary compositions. While a similar agreement in refractory composition between the host star and exoplanet and a considerable depletion of volatiles in the planetesimals might be expected, it remains unconfirmed and rigorous testing is essential. The bulk composition of exoplanetary material, a puzzle WDs help solve, offers a crucial piece of this testing process. However, WDs alone cannot provide the host star's composition. To address this, binary star systems present a unique opportunity. Just as planets and stars, binary systems form out of the collapse of the same cloud of interstellar gas and have similar ages and initial compositions, especially if the components of the system evolve independently, as is the case for wide binaries. Such stars have been found to be more chemically homogeneous than random stars in the field \citep{Andrews2019widebinaries, Hawkins2020, EspinozaRojas2021}. Although other theories for the formation of these systems have been suggested, the components of wide binaries are still expected to have similar ages and elemental abundances.

Examining wide binary systems with a metal polluted white dwarf and a main sequence (MS) companion unveils a promising avenue to utilize the shared origins of stellar siblings and their planets in deciphering the mysteries of exoplanet formation. The initial conditions of the system are determined by analyzing the spectra of the wide binary companion, still on the MS, and the composition of the exoplanetary material can be inferred from the spectrum of the metal polluted WD. In \citet{bonsor2021hoststar}, a pilot study is presented, using optical high-resolution spectra for the K-dwarf G200-40, a wide binary companion to the metal polluted white dwarf WD~1425+540. Abundances of C, N, O, Mg, Si, S, Ca, Fe, and Ni for the white dwarf suggested the ingestion of a rocky body rich in volatile elements \citep{Xu2017}.  \citet{bonsor2021hoststar} measured the chemical abundances of the K-companion and found that the refractory composition of the companion matched with the bulk exoplanetary material within the observational errors. However, C/Fe appeared to be depleted in the white dwarf planetary material compared to its host star. This is consistent with volatile depletion in planetesimals.
Throughout this work, we use different notations for abundances. The notation $\mathrm{X_1/X_2}$ refers to the logarithmic number abundances of element $\mathrm{X_1}$ relative to element $\mathrm{X_2}$. In contrast, the notation $\mathrm{[X_1/X_2]}$ represents abundances with respect to solar values, defined as $\mathrm{[X_1/X_2]=\log(X_1/X_2)-\log(X_1/X_2)_\odot}$, where $\mathrm{\log(X_1/X_2)_\odot}$ is the solar abundance ratio.

Each new system under examination offers a unique window into planet formation and the geological processing of planetary material (e.g., volatile depletion), depending on the elements found in the white dwarf spectrum. In this study, we present a comprehensive chemical analysis of the polluted white dwarf SDSS J082019.49+253035.3 (WD J0820 hereafter) and its binary companion HD69962, which is on the main sequence. This binary system, initially identified as a wide binary with main-sequence and white dwarf components by \citet{ElbadryRix2018}, has been further confirmed by the updated catalog of \citet{Elbadry2021}, specifying a projected separation of approximately $\sim7980$ au.

Section \ref{sec:observations} outlines our schematic representation of the evolution of the system, along with observations for both components, and spectroscopic analysis. In particular, subsection \ref{sec:paramsabund} presents the stellar parameters and abundances of the MS star while subsection \ref{sec:wdsparams} presents the abundances derived from the white dwarf spectra. Section \ref{sec:results} shows the determination of the accreted planetary material composition and the comparison between the abundances of the star and the planetesimal, assuming the starting composition of the system is given by the binary companion. Section \ref{sec:discussion} explores various scenarios to interpret the abundances observed in the white dwarf, accounting for sinking effects, and we summarize our conclusions in Section \ref{sec:conclusions}.

\section{Concepts,  data and methods}\label{sec:observations}

\begin{figure*}
\centering
\includegraphics[width=0.78\textwidth]{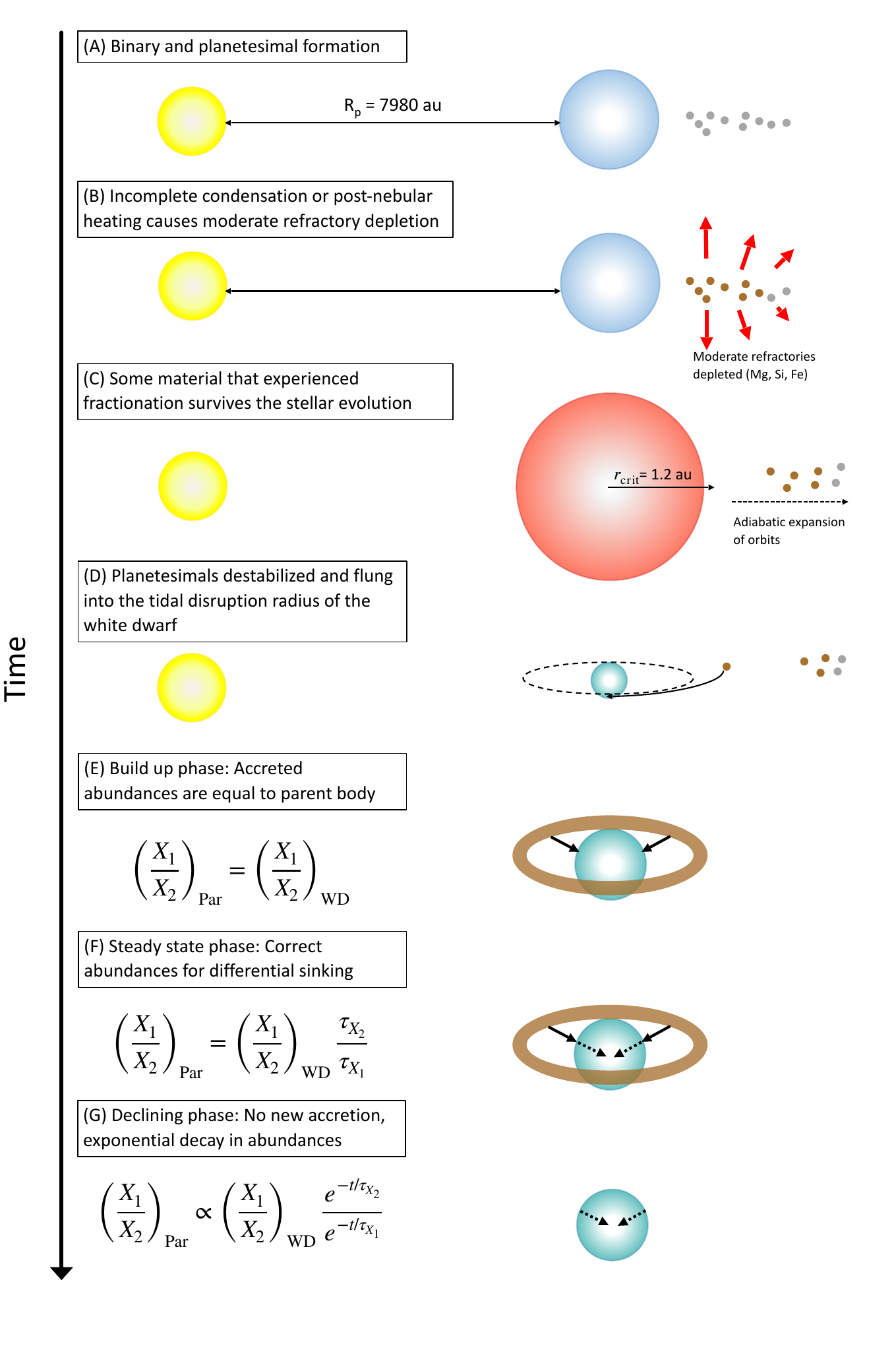}
  \caption{Formation of the binary and hypothesised evolution of the binary components and planetary material. $R_{\mathrm{p}}$ is the projected separation between the binary components, and $r_{\mathrm{crit}}$ is the critical initial semi major axis that a planetesimal must have in order to survive through the stellar evolution to the white dwarf phase assuming an eccentricity of 0. The depletion of moderate refractory relative to refractory elements in (B), indicated by red arrows, can occur both during and after nebula condensation. The different colors of the planetesimals represent whether they have experienced depletion of mildly refractory elements (brown) or not (gray). The white dwarf is being observed in declining phase (G). Equations in phases (E) to (G) show how the relative abundances of elements $\mathrm{X_1}$ and $\mathrm{X_2}$ change from the parent body depending on the accretion phase. }
     \label{fig:diagram}
\end{figure*}

\subsection{Stellar and planetary evolution}

A schematic representation of our binary system's evolution is illustrated in Figure~\ref{fig:diagram}. The beginning of this system lies in the two components of the wide binary, originating from the same primordial material and sharing an identical initial composition (A), which also serves as the building blocks for the planetesimals in this system. Throughout the planet formation process, various mechanisms can act to modify the abundance of rocky material within the system (e.g., B). In this work, we will use the term 'element fractionation,' referring to the concentration of elements in the solid phase compared to the gas phase. This term encompasses both incomplete condensation during the cooling of the protoplanetary disc and heating of material in the post-nebular phase. As the host star evolves into a white dwarf, it accretes a fraction of this rocky material that survived stellar evolution (C and D). However, as discussed later (Section \ref{sec:results}), the photospheric abundance of the resulting white dwarf may not precisely mirror the composition of the accreted planetary material (E-G).

\subsection{Observations}
\subsubsection{The main sequence star HD69962}

We obtained high-resolution, high signal-to-noise (SNR$\sim 70$) spectra for HD69962 with the Magellan Inamori Kyocera Echelle \citep[MIKE;][]{BernsteinMIKE2003} spectrograph on the 6.5m Magellan\ Clay Telescope, located at Las Campanas Observatory, over a run in November 2021. MIKE delivers full wavelength coverage from $\sim3500$ – $5000$ \AA\ in the blue arm and $\sim4500$ – $9000$ \AA\ in the red. The spectra were obtained with the $0.5''\times5.00''$ slit and $1\times1$ binning, which gives a typical spectral resolution R=$\mathrm{\lambda/\Delta \lambda}$ of $\sim55,000$ and $\sim45,000$ in the blue and red sides respectively. The data were then reduced using the CarPy MIKE pipeline \citep{Kelson2000CarPy, Kelson2003CarPy}. We then merge the different echelle orders, normalize the final spectrum using high-order spline fits, and correct to rest-frame using a cross-correlation function with an F-star template. We note that HD69962 also was observed and analyzed as part of the 16th data release of APOGEE \citep{Jonsson2020APOGEE}. This is an instrument that covers the infrared region, from 1.51 to 1.70 $\mu$m, and has a resolving power of approx. 20,000. We do not analyze this spectrum here but we consider the values reported for comparison with our analysis.

\subsubsection{The white dwarf SDSS J082019.49+253035.3}

The Sloan Digital Sky Survey (SDSS) observed SDSS\,J082019.49+253035.3 on 23rd November 2011. The SDSS DR10 spectrum was reported in \citet{Kepler2015sdss, GentileFusillo2015sdss} where it was identified as a DBAZ white dwarf{, that is, a white dwarf with a helium dominated atmosphere, also showing hydrogen and heavy elements in its spectra}. SDSS has a resolution R\,$\sim$\,1800 and covers the wavelength region from 3550 to 10300\,\AA~with a SNR of 40 around the calcium H and K lines (3934 and 3968\,\AA).

\begin{table*}[ht]
    \caption{Parameters and abundances for HD69962.}
    \label{tab:HD69962params}
    \centering
    \begin{tabular}{l c c c c}
    \hline \hline
    - & Optical & APO1 & APO2 & APO3 \\
    \hline
    \teff & 6521 & 6164 & 6202 & 6402 \\
    \logg & 3.86 & 3.69 & 3.72 & 3.81 \\
    \feh & $-0.189\pm0.050$ & $-0.219\pm0.032$ & $-0.233\pm0.035$ & $-0.079\pm0.017$ \\
    $v\sin i$ & 63.9 & 81.3 & 81.4 & 82.1 \\
    $\mathrm{[Mg/Fe]}$ & $0.127\pm0.100$ & $0.241\pm0.024$ & $0.216\pm0.025$ & $0.354\pm0.019$ \\
    $\mathrm{[Si/Fe]}$ & $0.132\pm0.100$ & $0.119\pm0.027$ & $0.117\pm0.028$ & $0.163\pm0.017$ \\
    $\mathrm{[Ti/Fe]}$ & $0.193\pm0.090$ & $-0.483\pm0.079$ & $0.089\pm0.084$ & - \\
    $\mathrm{[Ca/Fe]}$ & $0.137\pm0.070$ & $-0.072\pm0.035$ & $0.132\pm0.037$ & $-0.179\pm0.031$ \\
    $\mathrm{[Cr/Fe]}$ & $0.188\pm0.090$ & $0.196\pm0.059$ & $0.145\pm0.061$ & - \\
    $\mathrm{[Na/Fe]}$ & $0.235\pm0.08$ & $-1.575\pm0.063$ & $-0.920\pm0.064$ & - \\
    $\mathrm{[Ni/Fe]}$ & $0.467\pm0.19$ & $0.020\pm0.040$ &  $0.141\pm0.043$ & $0.067\pm0.030$\\
    \hline

    \end{tabular}

\raggedright{Note: This table includes data from all 3 APOGEE observations and Optical MIKE measurements. Associated uncertainties for chemical abundances are presented alongside each measurement. Notice that the APOGEE observations have some flags reported in the text. The uncertainties for our optical measurements are $u_{\mathrm{T_{eff}}} = 150 $ K, $u_{\mathrm{\log g}} = 0.10$ dex; $u_{v\sin i} = 7$ km/s. Typical uncertainties reported for APOGEE are $u_{\mathrm{T_{eff}}} \sim 140 $ K and $u_{\mathrm{\log g}} \sim 0.08$ dex.}

\end{table*}

\subsection{Spectroscopic analysis of HD69962} \label{sec:paramsabund}

The spectral analysis of the main sequence star HD69962 was done using {\tt iSpec} \citep{BlancoCuaresma2014ispec, BlancoCuaresma2019ispec}. This framework is a python wrapper which includes several radiative-transfer codes, atmospheric models and line lists to measure stellar parameters and abundances with either the synthetic spectral fitting on-the-fly technique or the equivalent width method. In this work, we used spectral synthesis where the code minimizes $\chi^2$ between a synthetic spectrum and observations. To do that, we adopted the MARCS grid \citep{Gustafsson2008MARCS} with the Solar abundance scale of \citet{Grevesse2007solar} and the code MOOG \citep[2017 version,][]{Sneden1973MOOG}, considering local thermodynamic equilibrium.

\begin{figure}
\centering
\includegraphics[width=\hsize]{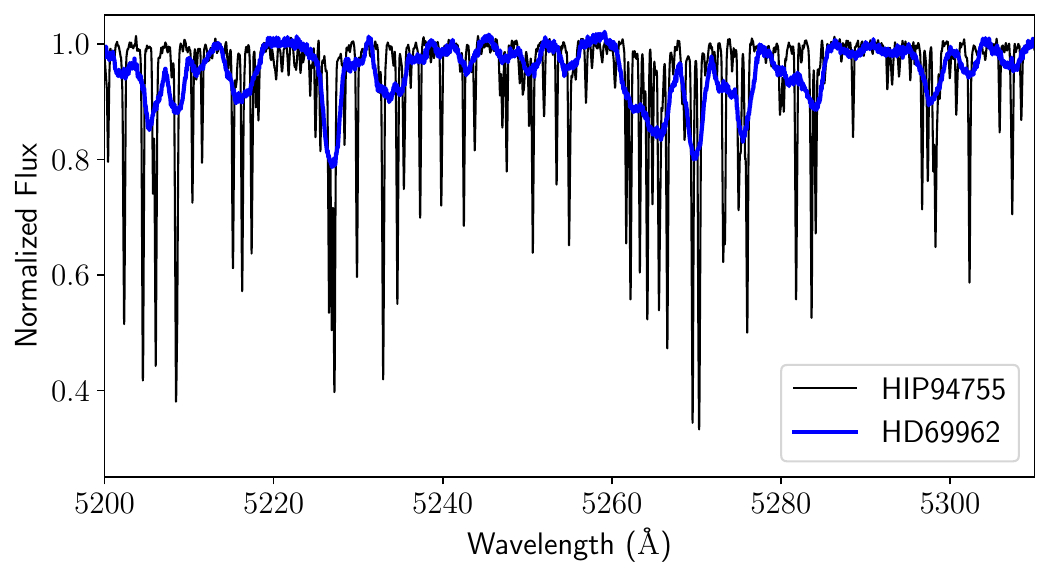}
  \caption{Region of the HD69962 spectrum compared to the Gaia Benchmark Star HIP94755, with similar atmospheric parameters. HD69962 shows the typical rotational broadening of the lines due to its fast rotation.}
     \label{fig:star}
\end{figure}

Figure \ref{fig:star} presents an example region of the stellar spectrum of HD69962 in blue, showing substantial rotational broadening of the lines (for a more detailed discussion, see Sect \ref{sec:ms_uncer}). In black, we include a spectrum from the Gaia Benchmark Star sample, version 3 \citep{Soubiran24}, HIP94755, for comparison. The Gaia Benchmark Stars have stellar parameters that are derived using fundamental methods and serve as reference stars beyond the Sun for spectroscopic studies \citep[see also][]{JofreGBS, HeiterGBS}. HIP94755 has similar parameters to HD69962 \citep{Soubiran24}, but its rotation is slow. It is due to the rotational broadening that we decided to use the synthesis method with $v \sin i$ as a free parameter to measure the atmospheric parameters and stellar rotation self-consistently. 
In {\tt iSpec}, used to fit the spectral lines, resolution, macroturbulence, and projected rotational velocity $v\sin i$ are degenerate parameters, all contributing to the broadening of the lines. As a result, we are unable to fully disentangle these effects. Given that our primary goal was not to obtain precise values for  $v\sin i$, but rather to find a consistent line broadening that allowed for the reliable determination of atmospheric parameters and abundances, we did not make additional efforts to separate the individual contributions of these broadening effects.
We used the atomic data and line list assembled for the {\it Gaia}-ESO Survey \citep{Heiter2021linelist}. Table \ref{tab:HD69962params} presents the atmospheric parameters measured for the star. Additionally, we derived a mass estimate of $1.29\pm0.20\ \mathrm{M_\odot}$ through the interpolation of MIST (MESA Isochrones and Stellar Tracks) isochrones \citep{Choi2016MIST}. The typical uncertainties associated with these measurements are $150$ K for the effective temperature, $0.10$ dex for \logg, and approximately $7$ km/s for $v\sin i$. These substantial uncertainties underscore the challenges encountered during the fitting procedure of synthetic spectra for this star with broadened lines, produced by its rotation.

Given that the spectral synthesis relies on a least-squares method to determine the optimal set of atmospheric parameters producing a synthetic spectrum best matching the observations, the uncertainties for these parameters are calculated from the associated covariance matrix, which in turn depends on the flux errors of our spectrum \citep{BlancoCuaresma2014ispec}. To assess the impact of the choice of free variables on our atmospheric results, we re-derived the parameters using different combinations of inputs in {\tt iSpec}. These included treating microturbulence as a free parameter versus calculating it from empirical relations, and using either both macroturbulence velocity and $v\sin i$ as free parameters or only one as adjustable by the model. In all cases, the differences in the results remained within the range of the reported uncertainties.

\begin{figure}
\centering
\includegraphics[width=\hsize]{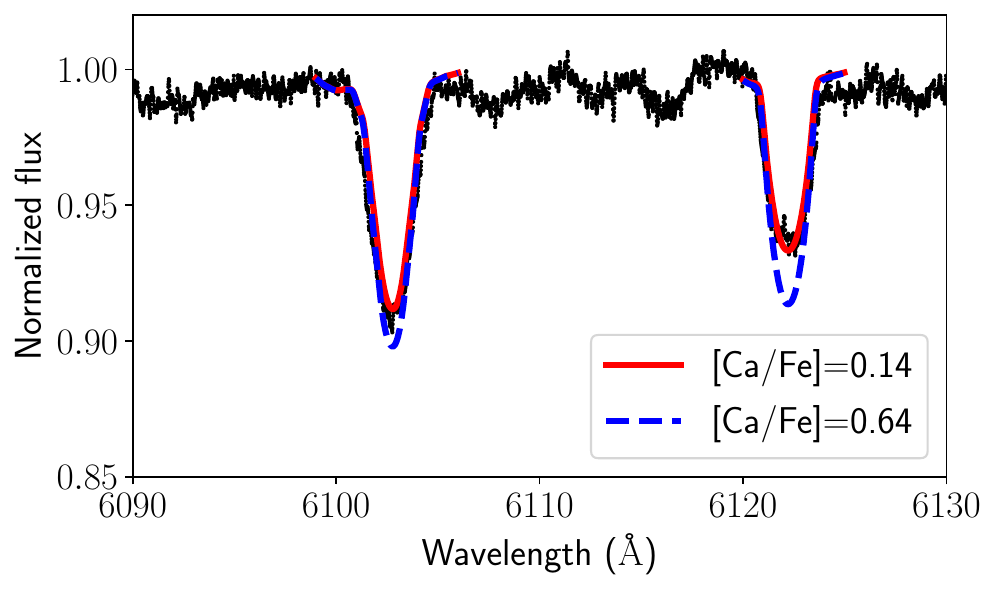}
  \caption{Sample fit of synthetic Ca lines at $6102.7$ and $6122.2$ \AA~(red solid line) to the observed spectrum of HD69962. A synthetic spectrum with 0.5 dex higher Ca is included (blue dashed line) to highlight the difference in abundance between the white dwarf and main sequence star.}
     \label{fig:Ca_star}
\end{figure}

Line-by-line abundances were determined using the {\it Gaia}-ESO line list as a reference for elemental spectral lines. However, we selectively retained lines based on visual inspection, focusing only on those present in the stellar spectrum. Lines with poor fits, where the synthetic profile deviates significantly from the observed line shape, leading to substantial errors, were filtered out. Subsequently, two iterations of a 2$\sigma$-clipping procedure were applied in each chemical abundance. In Table \ref{tab:lines} (Appendix \ref{app:Lines}), we present the final line list for the elements, including the parameters for each line. An illustrative example of Calcium lines, along with their corresponding fit, is depicted in Figure \ref{fig:Ca_star}.  We have included a synthetic spectrum with an abundance 0.5 dex higher than our best fit for the star, similar to the Ca abundance measured for the white dwarf (see Section~\ref{sec:wdsparams}), to show that such a high difference would be measurable from the spectrum, confirming the contrast in Ca between main sequence star and white dwarf.

Additionally, this star is included in the APOGEE-2 DR16 \citep{Jonsson2020APOGEE} and was analyzed using ASPCAP, the APOGEE Stellar Parameters and Abundances Pipeline.
There are three sets of observations for HD69962, with multiple visits each, to which we will refer as APO1, APO2, and APO3 from hereon. All of them have the flag SUSPECT\_BROAD\_LINES, confirming the broad absorption lines typical of a star with substantial rotation. In Table \ref{tab:HD69962params}, we include the values reported for all APOGEE observations. These correspond to the calibrated effective temperature, $\log g$ values, reported $v\sin i$, and abundances. A formal discussion on the calibration of parameters can be found in \citet{Jonsson2020APOGEE}. It might be worth mentioning that the uncalibrated values of \teff\ and \logg\ are also included in this Data Release and are closer to the values we measure from optical spectra.
The APO1 and APO2 observations have the ASPCAP flag COLORTE\_WARN, indicating a discrepancy in spectroscopic temperature of more than 500 K with the photometric temperature. Specific parameters can also present flags. For $v\sin i$, all observations have the GRIDEDGE\_WARN flag, indicating that the measured rotation rate is close to the edge of the grid.

In addition to the reported flags, Table \ref{tab:HD69962params} shows that there can be large discrepancies in the abundances from different APOGEE observations. Although these infrared observations allow us to check the consistency of the measured atmospheric parameters and abundances in general, assessing the accuracy of observations is complicated, and abundances derived from infrared and optical data do not always agree \citep{JofreIndustrial, Hegedus23}. Moreover, the flags and warnings in the APOGEE analysis and the differences between different observations confirm the difficulty in the derivation of parameters and abundances for this star. Thus, we have decided to include all measurements (3 from APOGEE and our optical determination from MIKE data) in our analysis and discussions.

\subsection{Spectroscopic analysis of SDSS\,J082019.49+253035.3} \label{sec:wdsparams}

\begin{figure}
\centering
\includegraphics[width=\hsize]{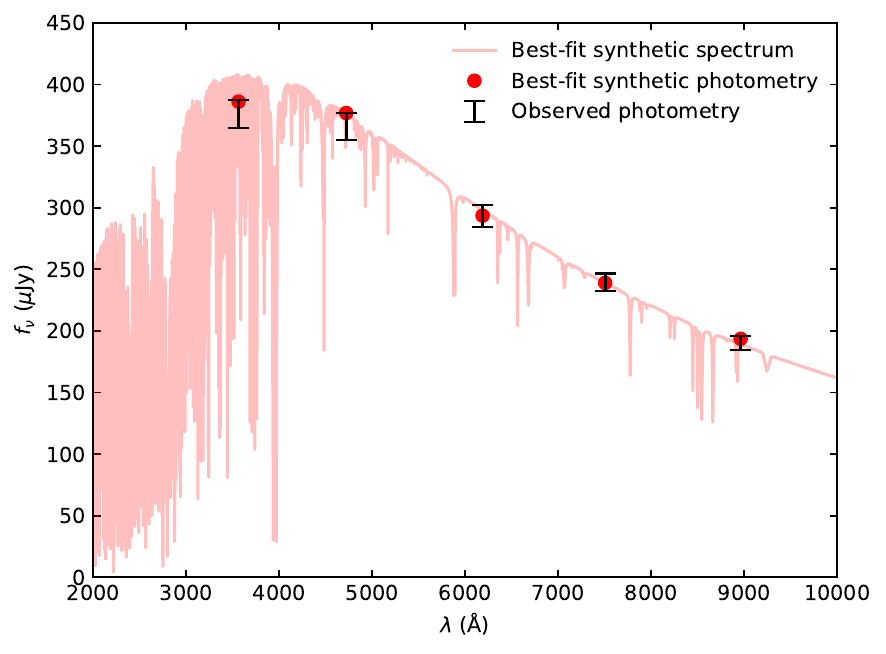}
  \caption{Spectral energy distribution showing the SDSS \textit{ugriz} photometry for SDSS\,J082019.49+253035.3 as black error bars, with the best fitting model photometry points as red circles.}
     \label{fig:SED-WD}
\end{figure}

To determine the white dwarf parameters and abundances of the planetary material accreted by WD\,J0820, a multi-dimensional grid of 1D model atmospheres with varying effective temperatures, surface gravities, and elemental compositions \citep{Dufour2007spectral,blouin2018new} is used to compare with photometric and spectroscopic data. Heavy elements and hydrogen in cooler polluted white dwarfs affect the pressure and temperature structure of the white dwarfs and so affect the derived white dwarf parameters \citep{dufour2012detailed,coutu2019analysis}. Therefore, an iterative procedure was used which alternates between fitting the white dwarf parameters (effective temperature and surface gravity) and the abundances of the metals \citep{Klein2021discovery}. The initial estimates of the white dwarf parameters were determined by using $\chi ^2$ minimization to fit the atmospheric models to broad-band SDSS \textit{ugriz} photometry, reported in Table \ref{tab:WD}, scaling to the distance derived from \textit{Gaia}. Figure~\ref{fig:SED-WD} shows the spectral energy distribution (SED) of the white dwarf alongisde with its best-fit spectrum (the same as in Figure \ref{fig:Models-all}). Once a best fitting solution is measured, the white dwarf parameters are fixed whilst the metal abundances are adjusted to fit the absorption lines of the metals in the SDSS spectrum. The photometric fit for the white dwarf parameters and the spectroscopic fit for the metal abundances are repeated until there is internal consistency between them. Fits to specific WD spectral lines can be found in Appendix \ref{app:WDfit}.

The final fitted white dwarf parameters and metal abundances are reported in Table \ref{tab:WD} and the model fit to the SED is shown in Figure~\ref{fig:SED-WD}. There are two key contributions to the abundance errors of the metals and these were calculated and added in quadrature, the first from the spread in abundances derived from different metal lines of the same element, and the second from the uncertainty in the effective temperature \citep{Klein2021discovery}. The derived white dwarf effective temperature is cooler than those derived in \citet{GenestBeaulieu2019photometric} (13294\,K using a spectroscopic method and 12692\,K using a photometric method), however, these differences are naturally explained by the omission of metals in their models. Upper limits of elements not measured in the spectrum were derived by paralleling the methods in \citet{LauraR2024} by finding an equivalent width upper limit that would have lead to a 3$\sigma$ detection for the strongest line for that element in the spectral range of SDSS. Upper limits were found for O, Al, Na, Cr, and Ni and are reported in Table \ref{tab:WD}.

   \begin{table}[h!]
      \caption[]{Properties of the white dwarf WD J0820 from \textit{Gaia} DR3 and measured abundances.}
         \label{tab:WD} 
         \begin{tabular}{p{0.35\linewidth}p{0.24\linewidth}|p{0.14\linewidth}}
            \hline
            \hline
            \noalign{\smallskip}
            Name:      &  \multicolumn{2}{l}{\textrm{SDSS\,J}082019.49+253035.3} \\
            Gaia source identifier: & \multicolumn{2}{l}{679451907694651904} \\
            ICRS RA (deg):  & \multicolumn{2}{l}{125.0812485217}  \\
            ICRS DEC (deg): & \multicolumn{2}{l}{25.5097129895} \\
            G (mag) & \multicolumn{2}{l}{17.621 (0.001)} \\
            \noalign{\smallskip}
            \hline
            \noalign{\smallskip}
            \teff ~ (K): & \multicolumn{2}{l}{11900 ~ (320)} \\
            \logg ~ $\mathrm{(cm\,s^{-2})}$: & \multicolumn{2}{l}{8.06 ~ (0.04)} \\ 
            $\mathrm{M_{WD}}$: & \multicolumn{2}{l}{0.619	~ (0.025)} \\
            $\log \mathrm{q^*}$: & \multicolumn{2}{l}{$-$4.99} \\
            \noalign{\smallskip}
            \hline
            \noalign{\smallskip}
            \textit{u} (mag) & \multicolumn{2}{l}{17.537 ~ (0.009)} \\
            \textit{g} (mag) & \multicolumn{2}{l}{17.521 ~ (0.005)} \\
            \textit{r} (mag) & \multicolumn{2}{l}{17.754 ~ (0.006)} \\
            \textit{i} (mag) & \multicolumn{2}{l}{17.950 ~ (0.008)} \\
            \textit{z} (mag) & \multicolumn{2}{l}{18.185 ~ (0.024)} \\
            \noalign{\smallskip}
            \hline
            \hline
            \noalign{\smallskip}
            Element $\mathrm{X_1}$ & $\mathrm{\log n(X_1)/n(He)}$ & $\mathrm{[X_1/Fe]}$\\
            \hline
            H & $-5.4\pm0.10$ & $-3.35$\\
            Ca & $-6.8\pm0.10$ & $0.95$\\
            Mg & $-6.3\pm0.14$ & $0.24$\\
            Fe & $-6.6\pm0.20$ & $0.00$\\
            Ti & $-8.6\pm0.20$ & $0.55$\\
            Si & $-6.2\pm0.20$ & $0.34$\\
            O & < $-5.0$ & < $0.39$\\
            Al & < $-5.9$& < $1.78$\\
            Na & < $-6.3$& < $1.58$\\
            Cr & < $-6.5$& < $1.91$\\
            Ni & < $-5.7$& < $2.12$\\ 

            \noalign{\smallskip}
            \hline
            \noalign{\smallskip}
            $\log \tau _{\mathrm{Ca}}$ (yr) & \multicolumn{2}{l}{6.18} \\ 
            $\log \tau _{\mathrm{Mg}}$ (yr) & \multicolumn{2}{l}{6.31} \\ 
            $\log \tau _{\mathrm{Fe}}$ (yr) & \multicolumn{2}{l}{6.05} \\ 
            $\log \tau _{\mathrm{Ti}}$ (yr) & \multicolumn{2}{l}{6.09} \\ 
            $\log \tau _{\mathrm{Si}}$ (yr) & \multicolumn{2}{l}{6.28} \\ 
            $\log \tau _{\mathrm{O}}$ (yr) & \multicolumn{2}{l}{6.43} \\ 
            $\log \tau _{\mathrm{Al}}$ (yr) & \multicolumn{2}{l}{6.27} \\ 
            $\log \tau _{\mathrm{Na}}$ (yr) & \multicolumn{2}{l}{6.31} \\   
            $\log \tau _{\mathrm{Cr}}$ (yr) & \multicolumn{2}{l}{6.07} \\ 
            $\log \tau _{\mathrm{Ni}}$ (yr) & \multicolumn{2}{l}{6.04} \\ 
        
            \noalign{\smallskip}
            \hline
            \noalign{\smallskip}
            
         \end{tabular}
     
     \textbf{Notes:} 
      RA and DEC use epoch 2016. Abundances measured for the white dwarf are presented both in log number abundances relative to He and as abundances relative to solar values. Diffusion timescales from \citet{koester2014frequency,koester2020new}.\\
     $^*$\,$\textrm{q}=log_{\textrm{10}}(M_{\textrm{CVZ}}/M_{\textrm{WD}})$, the ratio between the mass of the convective zone of the white dwarf, $M_{\textrm{CVZ}}$, to the total mass of the white dwarf, $M_{\textrm{WD}}$.
   \end{table}

\section{Results}\label{sec:results}

\subsection{Primordial material: The abundance of the stellar companion}

\begin{figure*}
\centering
\includegraphics[width=0.6\textwidth]{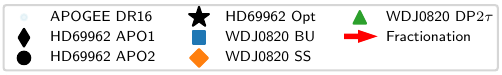}
\includegraphics[width=0.45\textwidth]{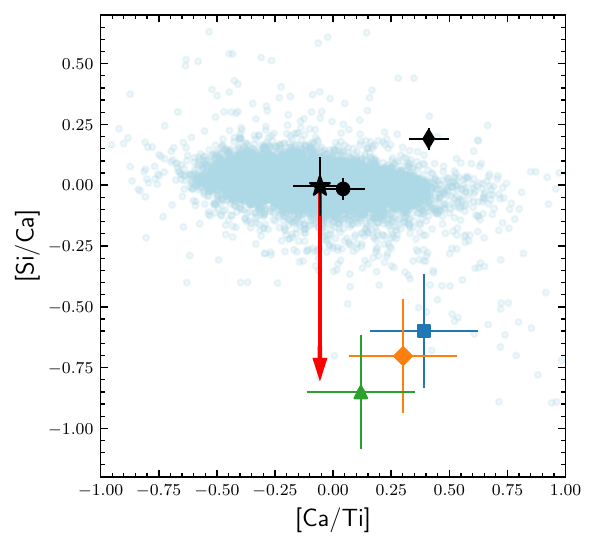}
\includegraphics[width=0.45\textwidth]{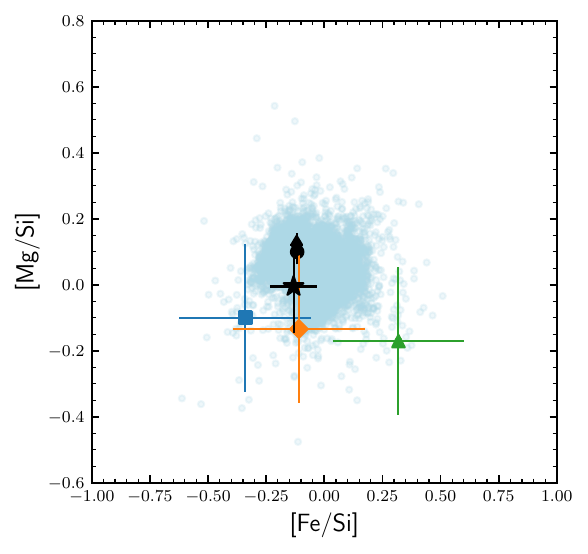}
\includegraphics[width=0.45\textwidth]{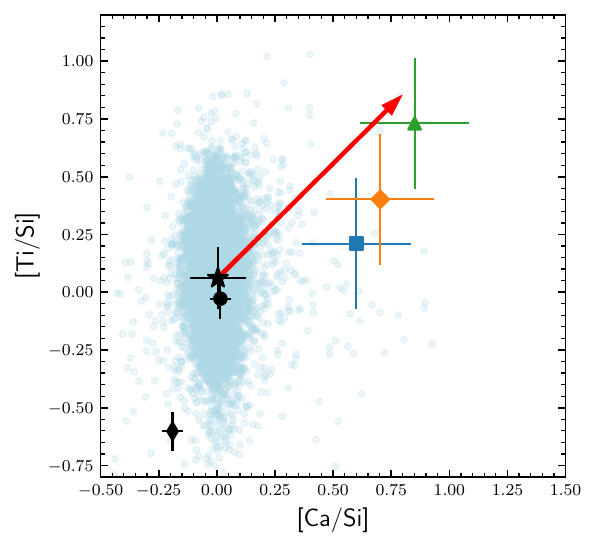}
\includegraphics[width=0.45\textwidth]{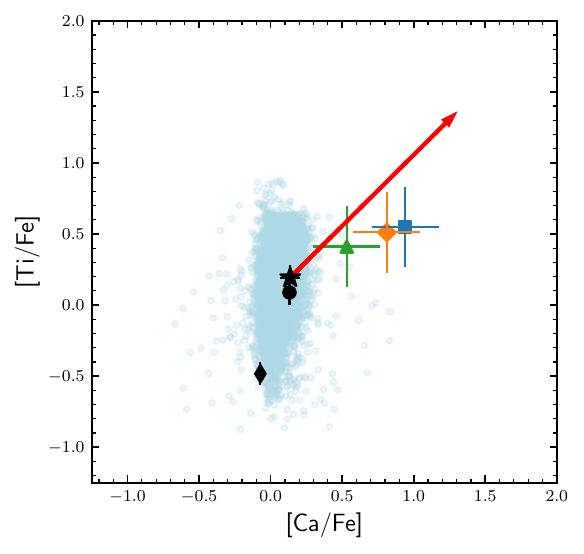}
   \caption{Abundance ratios of different elements. Black points represent two APOGEE measurements without flags and our optical spectra measurement of HD69962 (with $1\sigma$ uncertainties as errorbars), contrasted against other stars in APOGEE (light blue). Most of our measurements are consistent with the bulk of the Galaxy. The white dwarf WD J0820 abundances are depicted, considering potential phases of build-up (BU), steady state (SS) or declining phase after 2 sinking timescales of Mg (DP$2\tau$). A red arrow illustrates the potential effect of incomplete condensation and how it modifies the abundance ratios of the primordial stellar nebula, although heating in the post-nebular phase would produce a similar effect to these elements. The size of the red arrows reflects the predicted effect on elemental abundances when the material is heated to 1340 K.
   Considering fractionation, the abundances of star and accreted planetary material are consistent within uncertainties.}
              \label{fig:ratios}%
\end{figure*}

The chemical abundances of HD69962, detailed in Table \ref{tab:HD69962params}, are displayed in the abundance ratio plots depicted in Figure \ref{fig:ratios} alongside with those of Milky Way stars from the field, taken from APOGEE DR16 \citep{Jonsson2020APOGEE}, and the white dwarf companion. We exclude the measurements of the third APOGEE spectrum (APO3) here due to the absence of Ti abundance, which limit its utility in comparing with planetary composition. APO3 spectral analysis is indeed flagged as "Spectrum has significant number $(>20\%)$ of pixels in low persistence region: WARN (PERSIST\_LOW)". 

When focused on HD69962, we obtain consistent agreements for all abundances between our measurements and APOGEE, except [Ni/Fe]. For this element, we obtain 0.2 dex higher than  APOGEE, and we discuss it further in Sect.~\ref{sec:discussion}. Regarding Ti and Ca, our results agree with the measurements reported for APO2 only, and discussions about these measurements can also be found in Sect.~\ref{sec:discussion}. 
Overall, our measurements for the chemical composition of HD69962 align with other stars in the Galaxy, showing this star is a typical star in the solar neighborhood.

\subsection{The composition of planetary material from the abundances of the polluted white dwarf } \label{sec:WDmodels}

The abundance ratios of various elements observed in the atmosphere of a white dwarf may differ to those in the accreted body due to the differential sinking of elements in the white dwarf atmosphere. 

Polluted white dwarfs can be found and classified in three distinct phases of evolution, as depicted in Figure \ref{fig:diagram}, in phases from (E) to (G). In the build-up phase (E), accretion onto the white dwarf occurs without sufficient time for different elements to sink deeper into the atmosphere, remaining all visible in the atmosphere. Consequently, the measured abundances accurately reflect those of the engulfed material. Transitioning to the steady-state phase (F) in the evolution, the white dwarf continues to accrete material, but each element undergoes sinking at varying rates. 
During this phase, inferring the elemental ratio of the accreted planet requires scaling the measured abundance by the respective sinking timescales.
The final phase unfolds when the white dwarf ceases accretion (G). In this declining phase, elemental ratios decrease from the atmosphere exponentially, as each element progressively sinks without additional accretion.

As WD~J0820 shows no evidence for current accretion, from infrared or gaseous emission, it is not clear whether accretion recently started, or indeed has finished. From optical spectroscopy, there is no evidence of circumstellar gas in emission or absorption, nor is there any sign of excess infrared emission from dust. Although there could be an undetected disc of gas/dust, it is less likely that the white dwarf is actively accreting from a disc.

Relative sinking creates a characteristic fingerprint in relative abundances of various elements, which contrasts with the fingerprint of processes that may have occurred to the planetary body before it was accreted by the white dwarf. Here, we assume that the abundances currently in the atmosphere come from a single body, or smaller chunk of a larger body. This body formed out of material that had the same composition as the star, HD69962. The planetary body formed may have experienced incomplete condensation and/or lost moderately refractories compared to the star. These moderately refractory and volatile elements may be depleted during formation, due to impact heating post-formation, whilst on the main-sequence, or due to volatile loss during stellar evolution. The planetary body may also have undergone the large scale melting that leads to the formation of an iron core, with subsequent accretion of material predominantly from the core (or mantle) by the white dwarf. We use a Bayesian framework to assess the most likely combination of relative sinking, depletion of volatiles and segregation into the iron melt during core formation that best describe the observed abundances. 

This Bayesian model assesses the bulk chemistry of the planetary material accreted by the white dwarf, using the abundances of the main sequence star as a starting point. The effects of relative sinking are disentangled from those of planetary evolution by considering how multiple elements behave relative to each other. Distinct patterns in relative elemental abundances are associated with sinking, loss of volatiles, or the segregation of material into the iron melt that occurs during the formation of an iron core.
This approach allows us to distinguish the individual consequences of each step, with a particular emphasis on exploring the impact of planet formation on planetary composition. By disentangling the processes involved, we gain a better understanding of how each factor contributes to the final composition of the planetary material.

The Bayesian framework was developed in \citep{Harrison2018, harrison2021bayesian, Buchan2022} and can be found on GitHub\footnote{\url{https://github.com/andrewmbuchan4/PyllutedWD_Public}}. We edited the framework in order to avoid situational unphysical behavior associated with condensation of refractory elements at extreme temperature. We increased the minimum formation distance such that the temperature could not exceed 1700K ($\sim 0.1$ AU), and set the minimum possible Mg abundance to zero. For this work, unlike in previous studies where this had to be left as an unknown parameter, the initial abundances can be fixed to the abundances of the binary companion, HD69962. We use single values for the stellar abundances without considering the spread or sampling from the uncertainties. Although the errors in stellar composition are not negligible, they are considerably smaller than the uncertainties that would result from inferring the star's composition from the entire stellar population.

\begin{figure}
\centering
\includegraphics[width=0.42\textwidth]{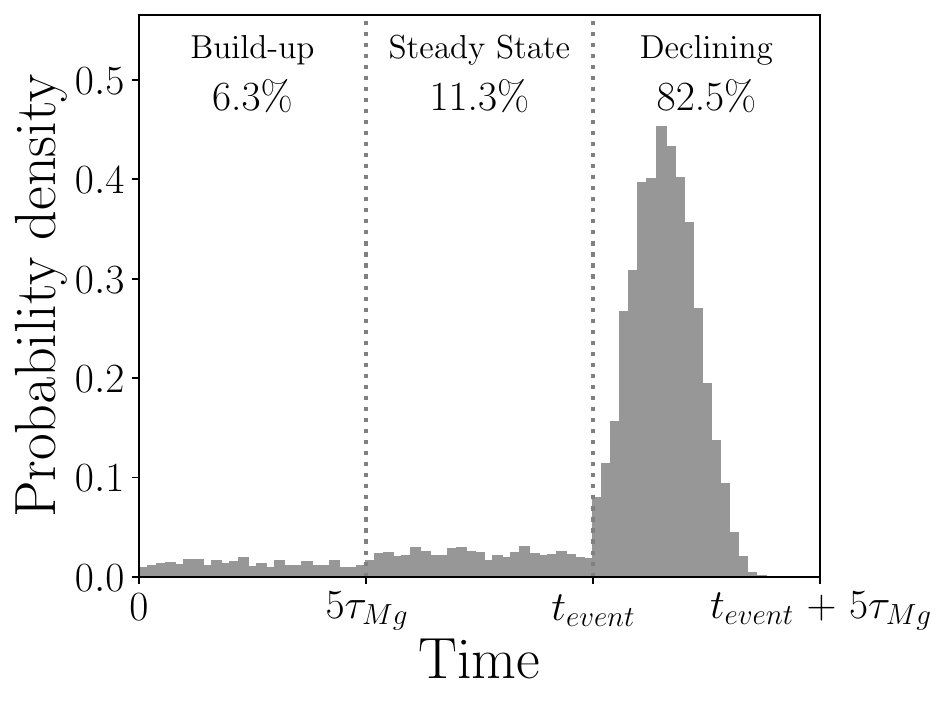}
\includegraphics[width=0.40\textwidth]{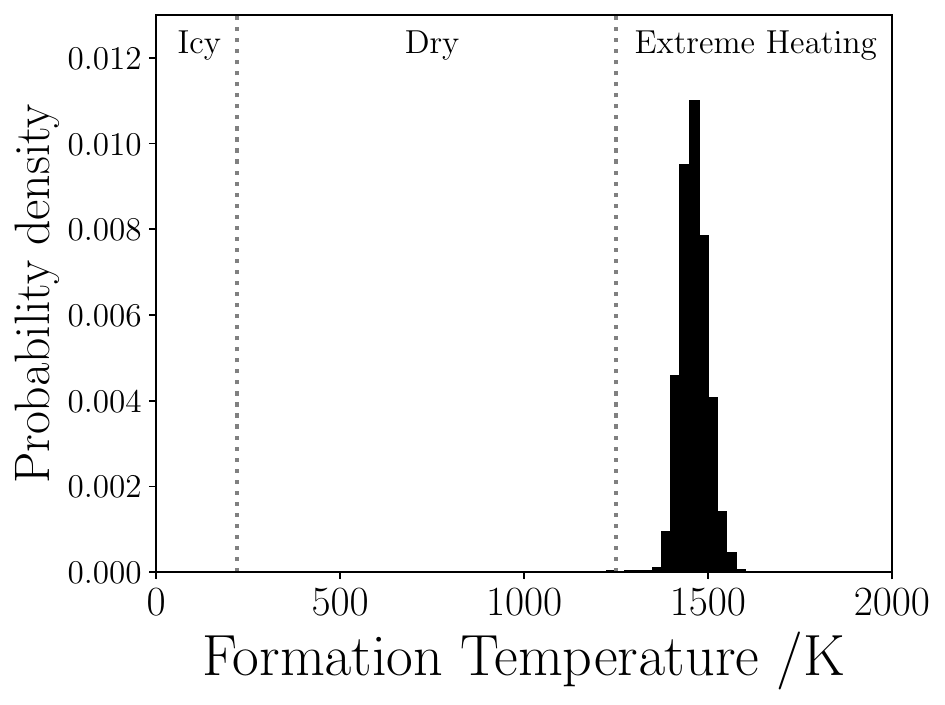}
  \caption{Posterior distributions of the time since accretion started onto the white dwarf (top panel) and the formation temperature of the planetary material needed to explain its observed composition (bottom panel). In the top panel $\tau_{Mg}$ is the Mg sinking timescale and $t_{event}$ represents the accretion event lifetime, indicating that the most probable explanation for the observed abundances is that the white dwarf is currently in the declining phase.}
     \label{fig:temptime}
\end{figure}

As a result of the Bayesian model applied, the most likely explanation for the observed abundances is accretion in the declining phase, involving material that is depleted in moderate refractories. This is illustrated in Figure \ref{fig:temptime}, where the median time since accretion finished is $2.78_{-3.98}^{+1.69}$ Myr, and the required temperature to explain the observed depletion is T = $1460_{-30}^{+40}$\;K. This interpretation of the abundances is discussed in more detail in the following sections.

\begin{figure}
\centering
\includegraphics[width=0.40\textwidth]{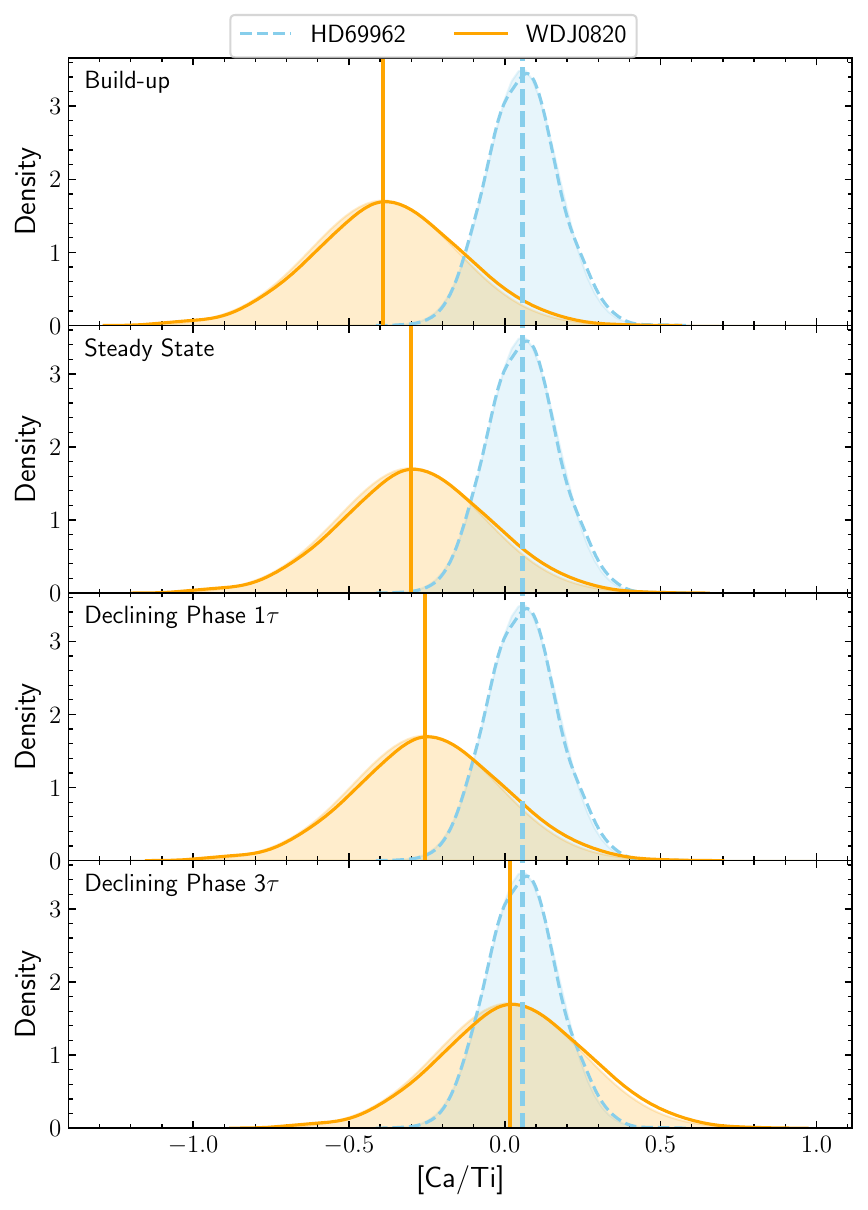}
   \caption{[Ca/Ti] abundance ratio distributions of the star HD69962 (blue dashed line) and its white dwarf companion WD J0820 (orange solid line). Different panels show the abundance of the white dwarf after being modified according to the accretion stage. From top to bottom we find build-up, steady state, declining phase after 1 and 3 sinking timescales of Mg ($\tau$)}
              \label{fig:cati}%
\end{figure}

\subsection{Sinking and WD accretion phase}
In deriving planetary material abundances for the precursor accreted body from the polluted white dwarf, it is crucial to consider the impact of sinking. Figure \ref{fig:ratios} provides a depiction of the planetary material abundance in three potential phases of white dwarf accretion: build-up (BU), steady-state (SS), and declining phase (DP) using as a reference the sinking timescale of Mg (See Table \ref{tab:WD}). In other words, these are the abundances measured in the white dwarf corrected by the effects of sinking.
We selected the Mg sinking timescale as a temporal reference for consistency, as elemental ratios are later expressed relative to Mg throughout the analysis. Although Fe is often used in stellar astronomy, it is less suitable here due to geological processes that could alter its composition in the planetesimal. Thus, Ca and Mg are considered better options, but Mg was ultimately chosen due to the relatively large Ca abundance observed in the white dwarf.

The key ratio for defining the accretion phase is predominantly [Ca/Ti]. Both Ca and Ti are refractory elements, the ratio of which is expected to remain unaffected by planet formation processes. Consequently, after correcting for the effects of sinking, the [Ca/Ti] ratio should be consistent with that measured for the star. The observation that [Ca/Ti] is higher in the white dwarf compared to the star implies that the most suitable phase to explain the ratio is DP, but SS cannot be ruled out given the uncertainties.

This can also be seen in Figure \ref{fig:cati}, which compares the [Ca/Ti] distribution of the binary companion HD69962 with that of the planetary material, assuming build-up, steady state, declining phase after 1 Mg sinking timescale, and declining phase after 3 Mg sinking timescales. 
Although the match between binary companion and planetary material composition is reasonably good for any accretion phase, considering the substantial uncertainties in the measurements, the distributions progressively converge, particularly in SS and DP. However, while the [Ca/Ti] ratio can provide insights into a specific accretion phase, a comprehensive assessment of the accretion phase requires considering other elemental ratios to determine the best fit that explains all the compositional data obtained for the white dwarf simultaneously.

\begin{figure}
\centering
\includegraphics[width=0.40\textwidth]{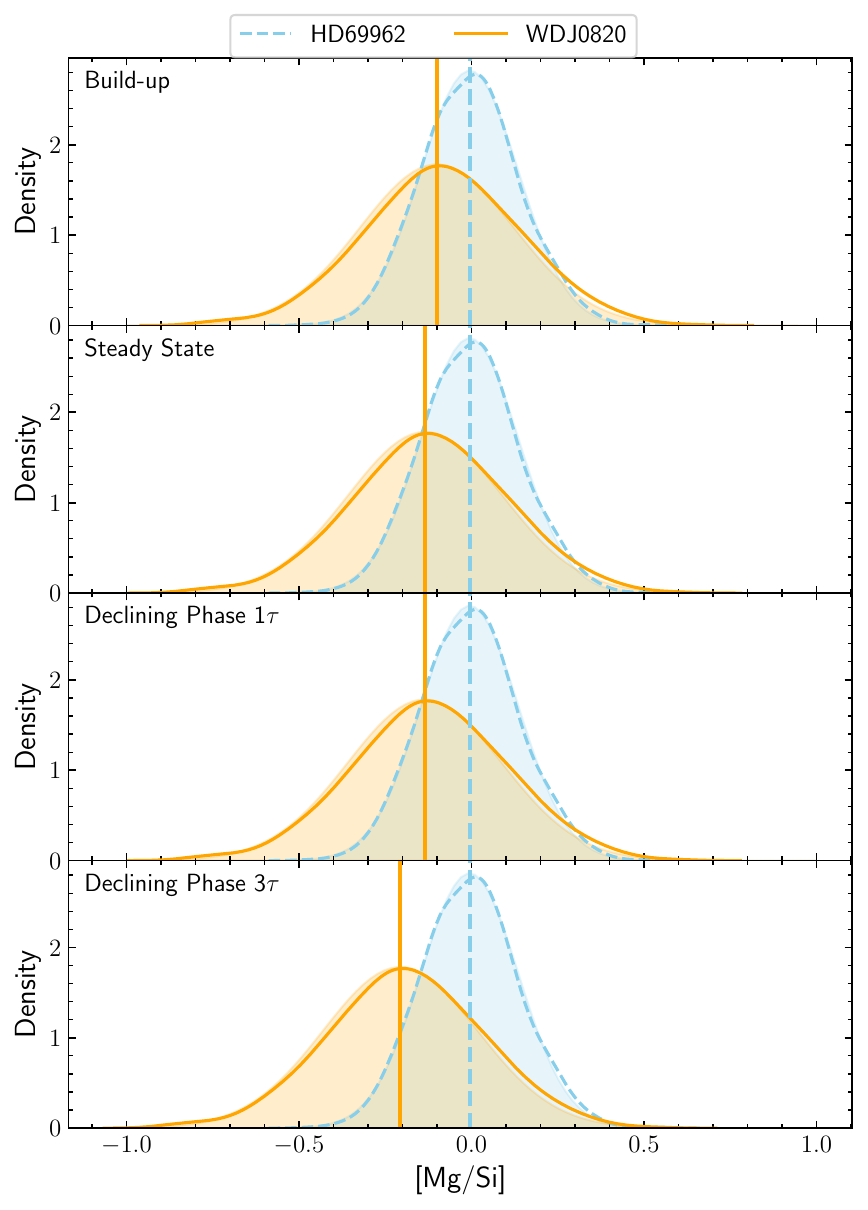}
   \caption{[Mg/Si] abundance ratio distributions of the star HD69962 (blue dashed line) and its white dwarf companion WD J0820 (orange solid line). Different panels show the abundance of the white dwarf after being modified according to the accretion stage. From top to bottom we find build-up, steady state, declining phase after 1 and 3 sinking timescales of Mg ($\tau$).}
              \label{fig:mgsi}%
\end{figure}

As seen in Table \ref{tab:WD}, Mg and Si exhibit similar sinking timescales, i.e., diffusion timescales, suggesting that the [Mg/Si] ratio should undergo minimal changes due to settling. This expectation aligns with the observed slight variation in the top-right panel of Figure \ref{fig:ratios}, where [Mg/Si] remains consistent between BU, SS, and DP, closely mirroring the [Mg/Si] ratio of the star within uncertainties. We notice that this ratio is closer to the abundances measured in the optical than APOGEE, but still consistent, regardless of the selected dataset.

Further confirmation of the lack of change in [Mg/Si] for the white dwarf accretion phase is illustrated in Figure \ref{fig:mgsi}. The distribution of the star and planetary material is consistent with every accretion phase alike, given that the observed changes are significantly smaller than those detected for the ratio of refractory elements.

\begin{figure}
\centering
\includegraphics[width=0.40\textwidth]{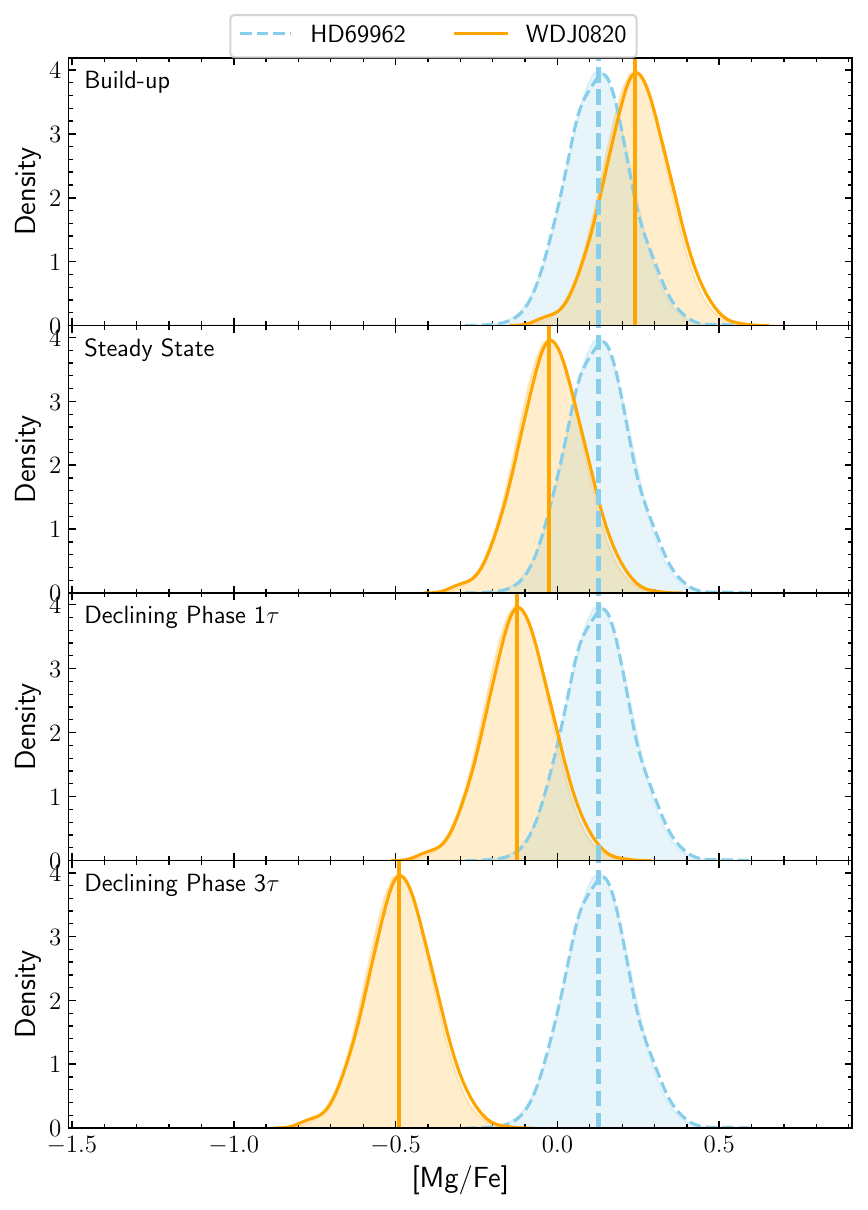}
   \caption{[Mg/Fe] abundance ratio distributions of the star HD69962 (blue dashed line) and its white dwarf companion WD J0820 (orange solid line). Different panels show the abundance of the white dwarf after being modified according to the accretion stage. From top to bottom we find build-up, steady state, declining phase after 1 and 3 sinking timescales of Mg ($\tau$)}
              \label{fig:mgfe}%
\end{figure}

\begin{figure}
\centering
\includegraphics[width=0.40\textwidth]{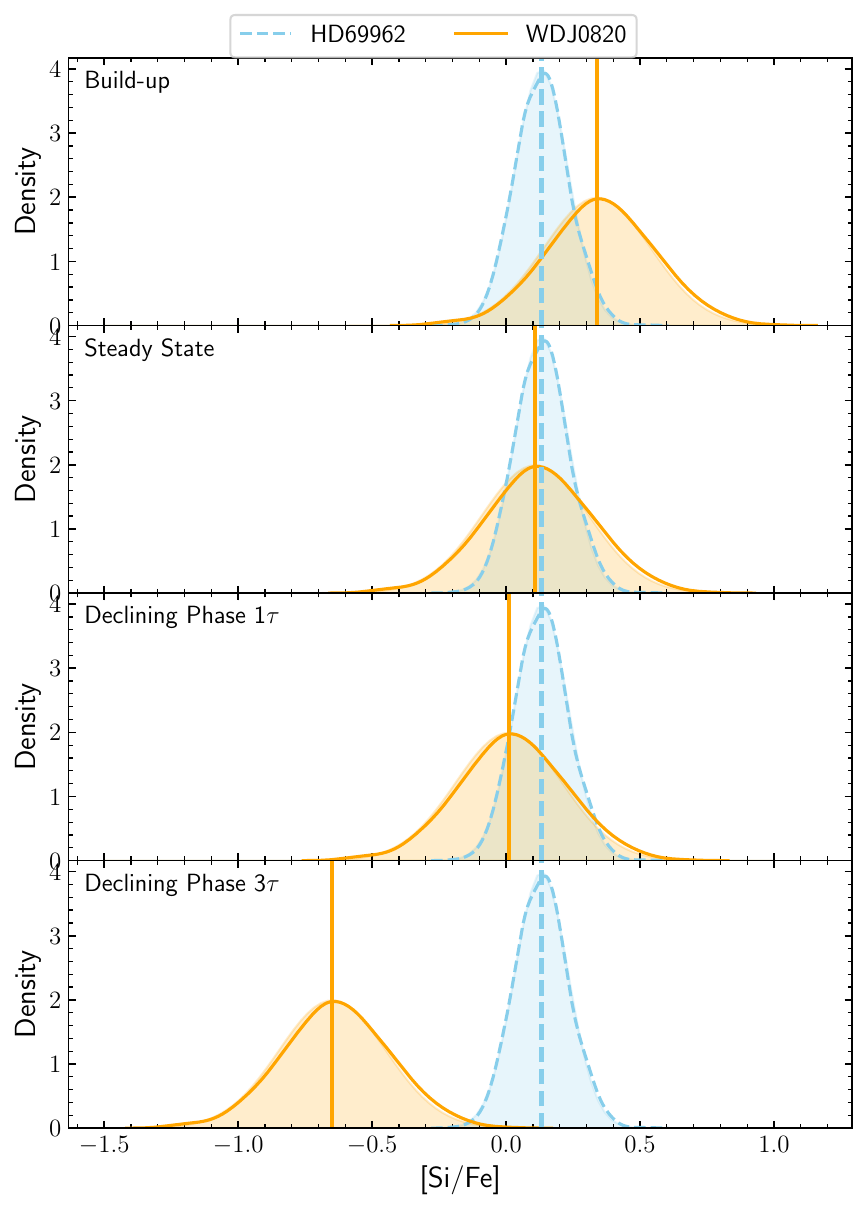}
   \caption{[Si/Fe] abundance ratio distributions of the star HD69962 (blue dashed line) and its white dwarf companion WD J0820 (orange solid line). Different panels show the abundance of the white dwarf after being modified according to the accretion stage. From top to bottom we find build-up, steady state, declining phase after 1 and 3 sinking timescales of Mg ($\tau$)}
              \label{fig:sife}%
\end{figure}

The role of Fe is key in this context due to its faster sinking rate compared to Mg and Si, despite sharing a similar condensation temperature \citep{Lodders2003}. Note that when estimating the parent body abundance from the white dwarf's current accretion phase, the difference in sinking rates (Fe sinking faster than Mg and Si) causes estimated abundance ratios such as [Mg/Fe], to decrease from BU to SS and then to DP, as shown in the equations in Figure \ref{fig:diagram}.

The planetary ratios [Mg/Fe] and [Si/Fe] experience substantial changes across different phases of white dwarf accretion, providing valuable insights into the system. Distributions of these ratios in the star, alongside corrected ratios for planetary material derived from the white dwarf abundance, are presented in Figures \ref{fig:mgfe} and \ref{fig:sife}. Analysis of these ratios strongly indicates that the optimal match occurs in the transition between the SS and the initial DP, excluding the possibility of an extended declining phase duration given that after accounting for sinking timescales, the abundance ratio of planetary material becomes notably smaller than that of the star, indicating inconsistency. The comprehensive analysis of all ratios collectively supports the conclusion that the accretion phase most consistent with the observed chemistry of the planetary material is the beginning of the declining phase.

An important conclusion can be deduced from these presented ratios.
Grouping elements based on their condensation temperature reveals a distinct pattern. Elements such as Mg, Si, and Fe, characterized by moderately refractory properties, share similar condensation temperatures. In contrast, Ca and Ti, deemed refractory elements, exhibit higher condensation temperatures. When grouping elements by condensation temperature, they consistently maintain the abundance of the star. In other words, instead of considering the abundance of each element independently, we analyze the ratios of elements with similar condensation temperatures. The relative abundances of these grouped elements are expected to be roughly the same as those in the star.
This observation shows a fundamental principle: any process influencing the planetary composition will produce a similar effect on elements with similar condensation temperatures. Therefore, when elements are categorized according to this characteristic, they consistently mirror the star's abundance. 

While the already discussed ratios ([Ca/Ti], [Mg/Si], and [Mg or Si/Fe]) demonstrate consistency, Figure \ref{fig:ratios} highlights the contrast in other element combinations. For instance, [Si/Ca] in the top-left panel, [Ti/Si] in the bottom-left panel, and [Ti/Fe] and [Ca/Fe] in the bottom right, all exhibit considerable differences from the stellar composition. This holds true even when considering only the previously determined accretion phase for the white dwarf (DP) and accounting for the large uncertainties in the abundances. These inconsistencies for certain ratios suggest that only considering sinking effects is insufficient to explain the chemical composition of the planetary material. Instead, it strongly implies that additional processes must be at play during the planet formation stage, affecting the abundances and introducing variations from the primordial composition inherited from the star.

\subsection{Effects of planet formation on the composition of planetary material}

We can focus on the elemental abundance ratios exhibiting disparities. For instance, the [Si/Ca] ratio  is notably higher in the star compared to the planetary material (indicated by the triangle symbol in Figure~\ref{fig:ratios}), whereas ratios such as [Ti/Si] or [Ca/Fe] are lower. Given that the accreted body is enriched in refractory elements and mirrors the abundance of the star, this discrepancy suggests a significant depletion of Si from the primordial composition. A similar analysis of other ratios reveals a consistent depletion of Si, Fe, and Mg from the stellar abundance, indicative of the loss of moderately refractory elements, consistent with a heating process, or depletion due to incomplete condensation.
In this section, we consider that planetary bodies may display varying elemental ratios due to the incomplete condensation of nebula gas during grain growth in the protoplanetary disc. 

The effect of incomplete condensation on chemical composition is shown in Figure \ref{fig:ratios} as a red arrow, originating from the measured stellar abundance. The size of this arrow, i.e., how much the primordial composition changes, depends on the extent of fractionation endured by the planetary material. As such, we can interpret its direction as a guide to anticipate the expected changes in abundance. Notably, the tip of the arrow corresponds to a temperature of the disc of approximately $\sim 1340$ K, similar to the temperature estimate from the Bayesian model in Sect. \ref{sec:WDmodels}. For all combinations of elements presented, the measured abundance for the planetary material (indicated by the triangle symbol) aligns consistently with the primordial abundance (represented by the star symbol) plus heating in the direction of the arrow. Although the effect of this arrow considers incomplete condensation, for the elements Mg, Si, Fe, Ti, and Ca, the impact should be similar to that of post-nebular volatilization. However, this similarity does not hold for volatile or moderately volatile elements such as Na and Mn, which exhibit different trends depending on whether they are affected before or after the dissipation of the nebula \citep{HarrisonSB2021}.

\begin{figure}
\centering
\includegraphics[width=0.48\textwidth]{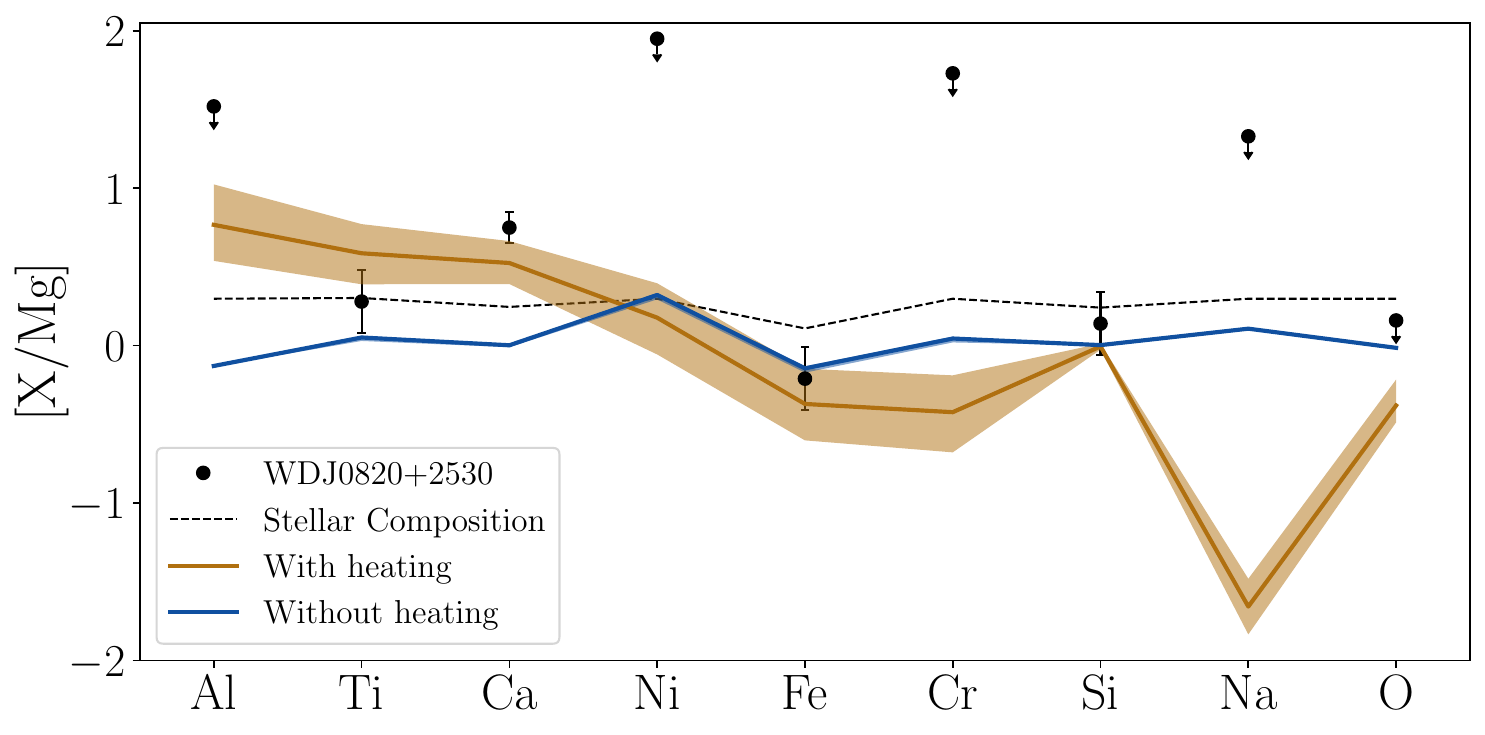}
  \caption{Comparison of abundances of WD and the main sequence star using optical data. Abundances are shown relative to Mg, and normalised to Solar composition on a log scale. The model with the highest Bayesian evidence, which includes heating, is shown in orange. This model also includes a feeding zone. The best model which does not include heating is shown in blue. This model cannot fit the data well, and so is heavily disfavoured. The small confidence interval for the non-heating model is due to its limited number of free parameters, mostly fixed or tightly constrained.} 
     \label{fig:WD}
\end{figure}

The influence of fractionation or heating is consistent with the optimal solution identified through the Bayesian framework, as depicted in Figure \ref{fig:WD}. This shows the measurements of the white dwarf (represented by points) and its stellar companion (illustrated by the dashed line), and fits that represent the posterior distribution derived from the Bayesian analysis, accounting for the effects of sinking. The model lacking heating (Figure \ref{fig:WD}, blue line) maintains consistency in [Fe/Mg] and [Si/Mg] ratios but inadequately explains the abundance Ti and Ca. Conversely, the model incorporating heating appropriately accounts for the elevated Ca abundance observed in the white dwarf while also capturing the variations in moderately refractory elements. The elements Al, Cr, Na and O would provide additional constraints on the model, but presently we only have upper limits for their abundances: these upper limits are consistent with the model result.

\section{Discussion}\label{sec:discussion}

In this section, we evaluate at what point in the system's evolution the exoplanetary material experienced elemental fractionation. We also consider alternative explanations for a possible heating source and discuss the assumptions involved in analyzing abundance data for main sequence star and white dwarf.

\subsection{Element fractionation during planet formation} \label{subsec:heating}

The first major stage of element fractionation occurs due to the incomplete condensation of gas as the protoplanetary disc cools.

Models, such as those proposed by \citet{Chambers2009models}, establish temperature and pressure conditions within the protoplanetary disc based on the characteristics of the host star. This enables the estimation of the fraction of an element in its solid state relative to its gaseous state at specific times and locations along the disc \citep{Harrison2018}. Such modeling can be tailored to any particular star and is consistent with the condensation temperatures reported for the Solar System, as derived from Solar photospheric abundances and meteorite measurements \citep[e.g.,][]{Lodders2003}. Accordingly, elements such as Ca and Ti are expected to condense at approximately 1400 K, whereas moderately refractory elements like Mg, Si, and Fe condense at around 1300 K. Consequently, based on our results from Section \ref{sec:results}, we can approximate the temperature of formation for the planetary material to lie between these two condensation temperatures.

To estimate at which point in the disc we find this temperature, we must first estimate the mass of the white dwarf progenitor. Since both stars in the binary system are coeval, the progenitor must have been more massive than HD69962 to reach the white dwarf stage. This sets a lower limit for the progenitor mass, indicating that the disc temperature exceeds that of a Solar-type star due to the stars’ initially higher mass. Using {\it wdwarfdate} \citep{Kiman2022wdwarfdate}, we derive the progenitor's mass based on the white dwarf's effective temperature and surface gravity, employing evolutionary sequences from \citet{Bedard2020models}, with thin outer hydrogen layers. Additionally, the code requires the use of an initial to final mass relation. We use relations provided by \citet{Cummings2018ifmr} or \citet{Marigo2020ifmr}, resulting in estimated progenitor masses ranging from $\sim1.4-1.5\,\mathrm{M_\odot}$ with uncertainties around $0.5\mathrm{M_\odot}$. Changing the metallicity of the system alters the estimated mass slightly, but always within the given uncertainties. This calculation suggests that the white dwarf progenitor is a late A type star, despite significant uncertainties. 

Assuming a white dwarf progenitor mass of $1.5-1.6,\mathrm{M_\odot}$ and using MESA evolutionary models, it's estimated that a planetary body must have an initial semi-major axis of of at least $\sim 1.0\,\mathrm{au}$
to survive stellar evolution. However, a planetary body at such distances would not reach the $\sim 1400\,\mathrm{K}$ necessary to explain the observed abundance ratios during its formation or the main sequence phase of its host star. To reach this temperature the body would need to be closer to the star, at $\lesssim 0.5\,\mathrm{au}$ initially, conflicting with the survival threshold of $\gtrsim 1.0\,\mathrm{au}$. This implies that either energetic collisions or radiogenic heating during the pre-main sequence or early main sequence phase, would be required to heat the planetary body if it is located at $1.0\,\mathrm{au}$ or more. Note that the distance from the star required to reach the necessary temperature could be extended outward when considering more complex structures for the disc and accounting for accretion heating of the protoplanetary disc \citep[e.g.,][]{Ueda2023}. 
While heating during the asymptotic giant branch phase could contribute, our analysis suggests it's unlikely to produce the observed elemental ratios (see Section \ref{subsec:AGB}). Alternatively, if the planetary body starts at $0.5\,\mathrm{au}$, it must migrate outward to a safe distance before the host star experiences thermal pulsations. 
These presented semi-major axis estimates are dependent on the model and formation time, but they provide useful references.

Although our estimate of the star mass is uncertain, even increasing it up to $3.5\,\mathrm{M_\odot}$ does not change our conclusion: Higher main-sequence temperature would push the location where the planetesimal experiences $1400\,\mathrm{K}$ farther away from the star, but since the maximum radius of the star during the AGB also increases with mass, the body would have to migrate beyond $2.0\,\mathrm{au}$ to survive. Thus, regardless of the star's mass, the challenge is the same: A planetary body in close initial proximity to the star that experiences enough heating, conflicts with survival thresholds that would allow it to be accreted during the white dwarf phase. 

In conclusion, elemental fractionation of the material polluting the white dwarf needed to happen early. Either the depletion of moderately refractories could have occurred during nebular condensation, followed by planetary body migration to a safer orbital distance to the star, or moderately refractory loss may have happened during the post-nebular phase via impacts and spontaneous isotope disintegration.

\subsection{Assumptions and limitations }

\subsubsection{Main sequence star abundance determination uncertainties} \label{sec:ms_uncer}

It it worth discussing the impact of our results on the uncertainties of the measured abundances in the main sequence star. The fact that HD69962 has substantial spectral line broadening produced by its rotation induces higher uncertainties in the abundances measured. The significant broadening, causes higher levels of blending with other lines, which can contaminate the abundance of some elements because of the absorption caused by the elements of neighboring lines. In broad lines it is also harder to accurately identify the level of the continuum. Nonetheless, as demonstrated by \cite{Casamiquela_fastrotator}, our method is still able to synthetize broad lines and determine consistent abundances for for stars rotating as fast as HD69962. After a careful revision of the line-by-line fits we selected only the lines that presented good fits. Additionally, we selected only the lines which have shown to yield trustable results in the literature.   

We can validate our results by comparing our abundances to those of APOGEE DR16, which provides independent measurements for three different spectra of the star. This is however an unfair comparison because APOGEE is focused on a different spectral range (thus other spectral lines) and the APOGEE pipeline has been optimized to work with red giants which rotate slow. 

Abundance ratio of [Si/Fe] is consistent among these 3 measurements and our own, despite the APOGEE Flag for that measurement about "Parameter in possibly unreliable range of calibration determination".  Our measurement of [Ti/Fe] of 0.193 agrees, within the uncertainties, only with one of the two measurements of APOGEE. While no flag is reported in APOGEE, the highly discrepant measurements hint to a very uncertain measurement, probably not recommended for validation with our result. Indeed, as concluded by \cite{Jonsson2020APOGEE}, Ti measurements should be treated with caution in APOGEE DR16. 

Regarding [Ca/Fe], our result agrees with only one from APOGEE. This measurement only reports a warning for the measurement of APO3, which has the flag "Parameter in possibly unreliable range of calibration determination". Considering distributions of abundances in the Milky Way, it is expected that alpha-capture elements have comparable levels of enhancement with respect to Fe in a star. The fact that Ca, Ti and Si are all between 0.1 and  0.2 dex is another hint toward the accuracy of our measurements. The difference of the Ca abundance between HD69962 and the white dwarf cannot be attributed to a wrong abundance measurement of the main sequence star.   

The abundance of [Mg/Fe], among the alpha-capture elements studied here, is the most difficult to determine. The two best lines in the optical for a large variety of stars are 613.8 and 613.9 nm \citep[see discusison in e.g.,][]{Jofre15, JofreIndustrial} and are blended in HD69962. The only line left with good fits presented is the 552.84 nm, which in \cite{Heiter2021linelist} was reported to be a good line for synthesis, but its atomic data quality was undetermined. The [Mg/Fe] abundance is obtained only from that line and while it is a bit lower than all three APOGEE values, is still within the uncertainties, and still consistent with Si, Ca and Ti. 

The measurements of [Cr/Fe] and [Na/Fe] agree well with APOGEE, but [Ni/Fe] disagrees by 0.2 dex, although our measurement has high uncertainties.  Considering that \cite{Jonsson2020APOGEE} concludes that Ni is among the best elements in APOGEE, and that our value of [Ni/Fe] of 0.4 is higher than the expected distributions of normal stars in the solar neighborhood, it is possible that our value is not correct. After removing lines with bad flags in the atomic data from \cite{Heiter2021linelist}, we used 7 lines for our measurement. Among these lines, only two of them (542.4 and 600.7 nm) have Y flags in \cite{Heiter2021linelist}, indicating they are good to be used to measure abundances with spectral synthesis, and the abundances of these two lines are of the order of $0.5$ dex.  We do not aim to understand the nature of this enhancement, because this value is still lower than the upper limit estimated to the white dwarf, making the abundances of this element not a key player in our conclusions. 

\subsubsection{Uncertainties in white dwarf measurements and atmospheric modeling}

As is the case of the main sequence star, measuring the element abundance from the white dwarf and then tracing it back to a possible planetary composition relies on a series of assumptions, that affect our results. 
One fundamental assumption is that the pollution of the white dwarf is produced through accretion from a single planetary body, although this may not hold true in all instances \citep{TurnerWyatt2020multiple,Johnson2022unusual}.

The scarcity of spectral lines in white dwarfs implies a significant challenge, not only because the measurement needs to be made with fewer spectral features but also due to potential uncertainties in the atomic data of these lines. 

On the other hand, transforming measured abundances into elemental ratios in the planetary material relies on the modeling of the white dwarf’s atmosphere and the settling timescales that can remove those observed heavy elements from sight. Given that white dwarfs in the temperature regime where we find WD J0820 have developed convective zones, the treatment of convection within the region and overshooting into the diffusion-dominated region below becomes crucial for understanding planetary material accretion. However, conventional 1D models, like those used in this study, apply the mixing length theory for convection and may underestimate the sinking timescales and the extent of mixing beyond the bottom of the convective layer, as exemplified by \citet{Cunningham2019} using 3D models of DA white dwarfs. This emphasizes the importance of incorporating a comprehensive treatment of white dwarf atmospheres in 3D, including effects such as overshooting, to mitigate potential biases in the measured planetary abundance ratios.

\subsection{Alternative explanations for the composition of the planetary material}

\subsubsection{Asymptotic Giant Branch heating} \label{subsec:AGB}

While moderately refractory element depletion can occur during the nebular phase due to incomplete condensation, it is also plausible for the planetary material to undergo compositional changes at later stages. In the post-nebular phase, radioactive decay and collisions may lead to melting and vaporization of species \citep{Siebert2018MnNa}, as mentioned in Section \ref{subsec:heating}. Alternatively, during the stellar evolution of the host star, particularly in the asymptotic giant branch phase, the planetary material may be subjected to heating by the star \citep{liyuqi2024thermalevol}.

Exoplanetary material engulfed by a white dwarf must endure the star's entire evolution, including dynamic interactions that may lead to migration, engulfment, or disruption. It is crucial to accurately model the stellar and orbital evolution to understand the survivability limits of planets. Moreover, stellar irradiation can mimic processes that happen during planet formation. As such, understanding the thermal evolution of a planet allows us to evaluate whether heating can melt the outer portion of the body, inducing a loss in moderately refractory elements even before reaching the white dwarf phase. Given that core-mantle differentiation and asynchronous accretion can affect elemental ratios such as Si/Fe or Ca/Fe \citep{Brouwers2023coremantle}, it is relevant to test if late-stage thermal evolution of the exoplanetary material could produce the signals we found in this system. Studies by \citet{liyuqi2024thermalevol} suggest that known pollutants found in helium-dominated white dwarfs are unlikely to be differentiated during the star's giant branch phases. An estimation using the specific parameters of this system reveals a similar conclusion.

Estimating the mass of planetary material needed to create the observed signature ($\sim3\times10^{21}$ kg) and assuming a density of 3-8 g/cm$^3$, we find that the accreted body's radius is about ~400-600 km. During the star's giant branch phases, the heating effects are significant but the duration is insufficient for substantial thermal penetration beyond the planetesimal's surface layer.  
The diffusion length scale remains too small to impact anything beyond the surface, as supported by numerical evidence \citep[e.g.,][]{liyuqi2024thermalevol}. Optimistically, a surface layer less than $\sim20$ km may undergo moderately refractory depletion. Thus, it's unlikely that the observed moderately refractory depletion in the white dwarf pollutant was caused by heating during the giant phases of stellar evolution.

\subsubsection{Crust accretion}

\begin{figure}
\centering
\includegraphics[width=0.5\textwidth]{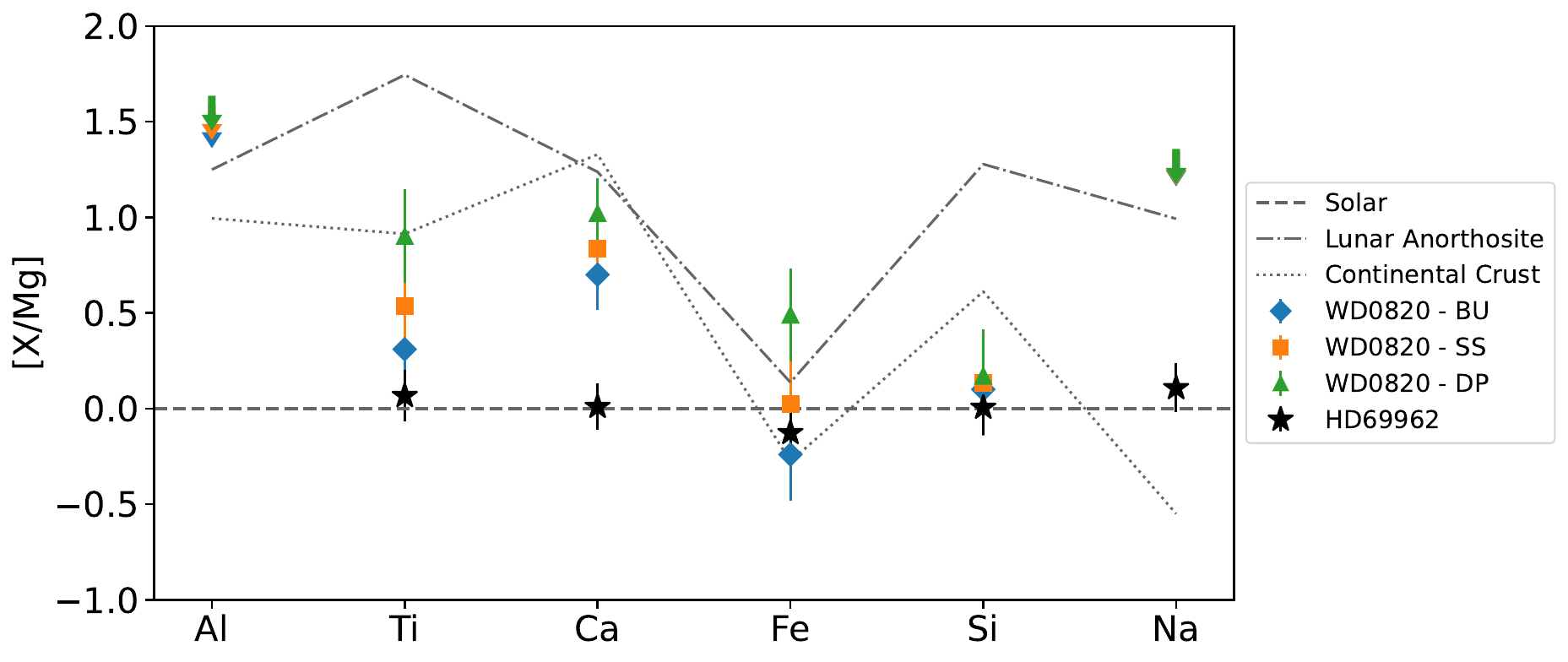}
   \caption{Comparison of the main sequence and white dwarf abundances in different accretion phases, including the typical composition of continental crustal material (dotted line) and Lunar anorthosite (dot-dashed lines). These abundances are normalized to Solar values. Downward arrows in this figure represent the upper limits measured in the white dwarf atmosphere.}
              \label{fig:alternatives}%
\end{figure}
An alternate explanation for the high Ca, Ti abundances of the planetary material accreted by WD J0820 is that the material present in the atmosphere currently is the crust of a planetary body. Crustal material has undergone further phases of successive melting and would be enhanced in elements such as Ca and Ti, but also Al. Na is a crucial element with the potential to distinguish between crustal material and material that has undergone successive heating. Unfortunately Na is not detected in the spectrum of WD J0820 and the Na upper limit does not distinguish between crust and loss of moderate volatiles. 

The abundances of WD J0820 were compared to both crustal material and Lunar anorthosite, a very Ca-rich rock (see Figure \ref{fig:alternatives}). However, the Ca/Si ratio of the white dwarf is not in agreement with the abundance of either material. Invoking the accretion of continental crust to explain the observed abundances would add extra complexity to the model, such as requiring selective accretion of the planetary material's surface rather than the entire body. Therefore, with the current available data, there is no reason to suggest that these are more likely explanations. The sodium abundance would act as a potential discriminator between crustal sodium-rich material and material depleted in moderate volatiles due to heating, which is sodium poor. The current upper limit on Na does not discriminate between the two scenarios.

\subsubsection{Primordial Cloud}

In this work, we have assumed that both stars in the binary system and the planet formed from the same cloud of gas and dust. However, the homogeneity of such clouds remains debated \citep[e.g.][]{Ramirez2019}. The discovery of a comoving pair with significant compositional differences, such as Kronos and Krios \citep{Oh2018,Miquelarena2024}, including important discrepancies in their metallicities, suggests that the original cloud might not have been homogeneous. Alternatively, other processes, such as planet engulfment \citep{Spina2021} or atomic diffusion \citep{Liu2021}, could have altered the abundance of one of the stars. \citet{Saffe2024} identified a wide binary pair of giant stars with a difference in composition of $0.073\pm0.035$ dex for volatile elements and $0.081\pm0.010$ dex for refractories, potentially indicating a primordial origin for such discrepancies. This finding implies that wide binaries might not always share identical abundances. Most wide binaries exhibit metallicity differences smaller than $\mathrm{\Delta[Fe/H]}=0.1$ dex between components \citep[e.g.][]{Behmard2023}. This difference falls within the error bars of our stellar composition measurements. Therefore, the depletion of mildly refractory elements observed in this study is unlikely to be explained by compositional differences between the F-star and the planet-forming material, though this possibility cannot be entirely excluded.

\section{Conclusions} \label{sec:conclusions}

To understand the interior composition of exoplanets, one common approach is to assume that planets share the composition of their host stars, as both originate from the collapse of the same primordial cloud. However, it is important to investigate how the planet formation process modifies the composition of the nebula. 

Spectroscopy of polluted white dwarfs provides the composition of the exoplanetary material they have accreted. When these white dwarfs are part of a wide binary system, they provide a unique opportunity to compare stellar and planetary compositions. Assuming chemical homogeneity within wide binaries, the companion composition serves as a proxy for the nebula's abundance, from which both stars and planets formed.

Following the work by \citet{bonsor2021hoststar}, who used one of these systems to confirm that the refractory composition of a planet and its host star match, we present the second paper in our series. In this study, we introduce a new wide binary system with a polluted white dwarf and a main sequence companion, exploring the link between host-star and planetary abundances.

We present detailed abundances of Fe, Mg, Si, Ca, and Ti for both white dwarf and its companion, a main-sequence F-type star. The companion is used to fix the initial conditions for the entire system, including the white dwarf’s progenitor. After considering the effects of sinking in the white dwarfs atmosphere, we derive the abundance of the accreted planetary material.

The key results of our work are summarized as follows:
\begin{itemize}
    \item Consistent initial composition: The abundances of the accreted planetary material, as seen in the white dwarf atmosphere are consistent with planet formation starting from the same initial abundances as its host star.
    \item Moderate refractories depletion: Before arrival onto the white dwarf the planetary material is depleted in moderate refractories, including Mg, Si and Fe, relative to refractory elements (Ca, Ti, Al). This depletion follows the relative volatility of elements, as characterised by their condensation temperatures. 
    \item Using condensation temperatures: Grouping elements by condensation temperature provides an effective diagnostic for linking planet-host star and planet composition. 
    \item Possible causes of depletion: The lack of moderately refractory species in the accreted planetary material by the white dwarf could have occurred due to incomplete condensation, heating during planet formation or subsequent migration, radiogenic heating, collisions or indeed during the evolution onto the white dwarf.
\end{itemize}

This work highlights the best means to utilise stellar compositions to inform our understanding of planetary compositions. Future observations of polluted white dwarfs in wide binary systems with multiple elements and different histories will improve the determination of interior compositions of rocky exoplanets and help understanding the intricacies of the planet formation process.

\begin{acknowledgements}
This work uses observations obtained through a Chilean Telescope Allocation Committee (CNTAC) proposal under ID CN2021B-56. We thank the referee for their valuable comments and suggestions, which have helped improve the clarity and quality of this work. C.A.G. acknowledges support from Agencia Nacional de Investigación y Desarrollo (ANID), Proyecto FONDECYT Iniciación 11230741. C.A.G and P.J. acknowledges support from ANID Proyecto FONDECYT Regular 1231057 and Millennium Nucleus ERIS  NCN2021\_017. LKR acknowledges support of an ESA Co-Sponsored Research Agreement No. 4000138341/22/NL/GLC/my = Tracing the Geology of Exoplanets. AB and LKR acknowledges the support of a Royal Society University Research Fellowship, URF\textbackslash R1\textbackslash 211421. AMB acknowledges the support of a Leverhulme Trust Grant (ID RPG-2020-366). S.X. is supported by the international Gemini Observatory, a program of NSF's NOIRLab, which is managed by the Association of Universities for Research in Astronomy (AURA) under a cooperative agreement with the National Science Foundation on behalf of the Gemini partnership of Argentina, Brazil, Canada, Chile, the Republic of Korea, and the United States of America. 

\end{acknowledgements}

\bibliographystyle{aa} 
\bibliography{wd}
\onecolumn
\begin{appendix}
\section{WD spectrum}
\label{app:WDfit}
\begin{figure*}[h!]
    \centering
         \includegraphics[width=0.49\textwidth]{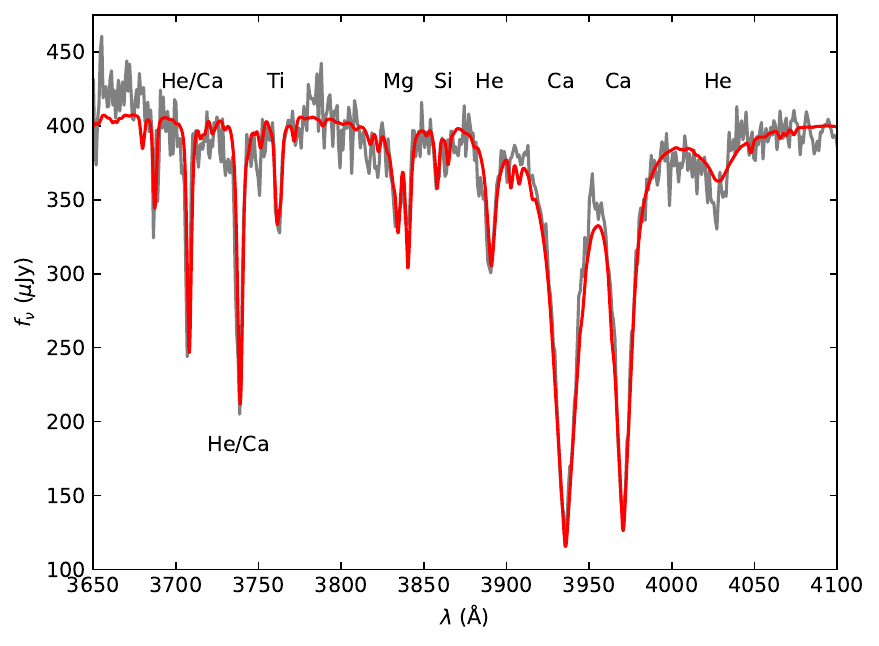}
         \includegraphics[width=0.49\textwidth]{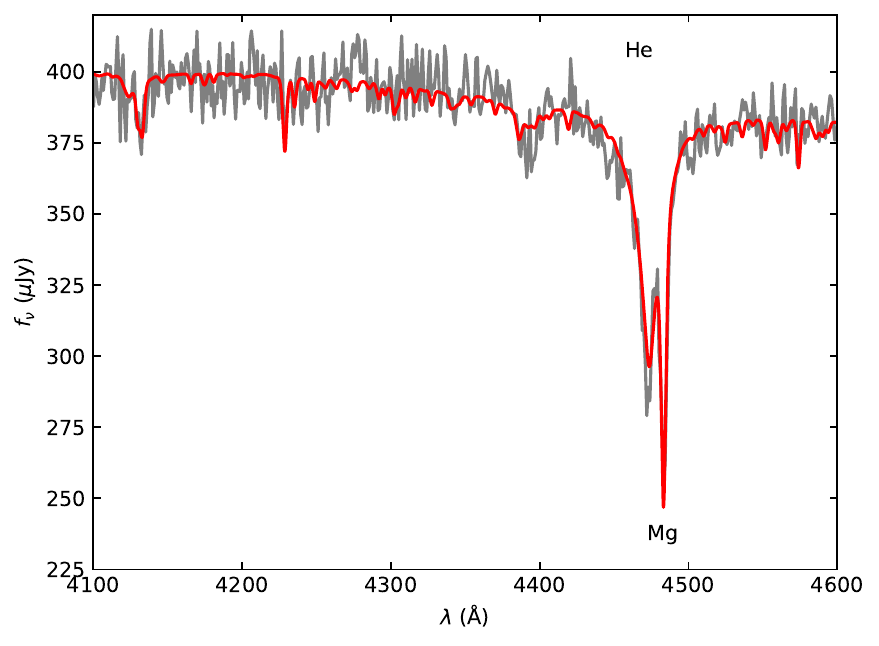}
         \includegraphics[width=0.49\textwidth]{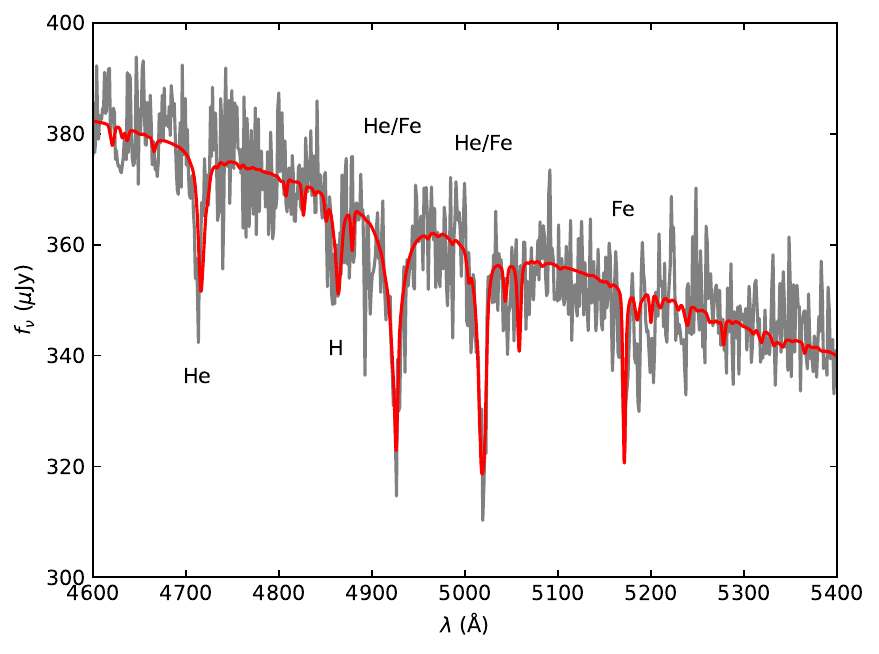}
         \includegraphics[width=0.49\textwidth]{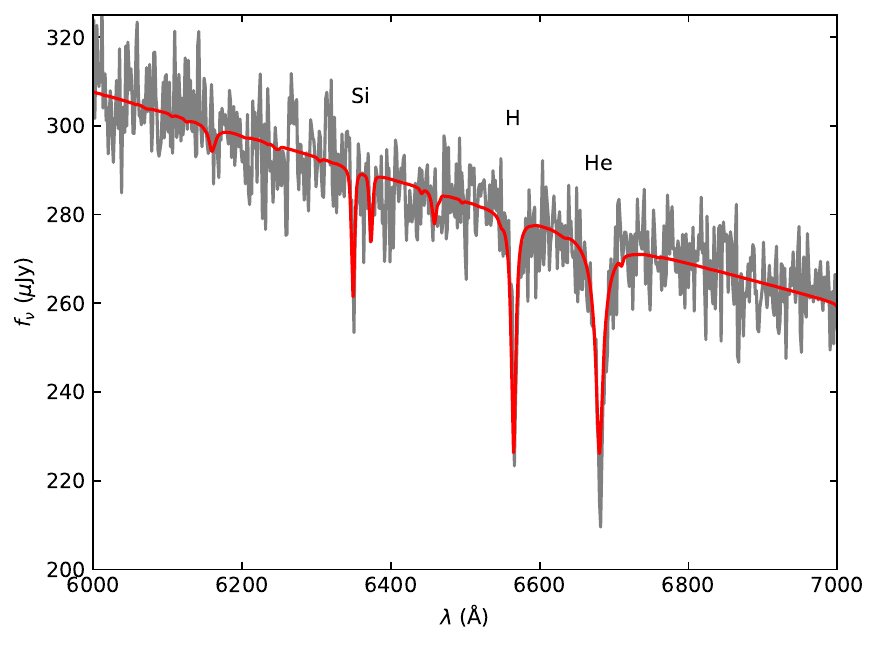}
        \caption{Four regions of the SDSS spectrum showing the best fitting atmospheric model to the data. Key lines have been labeled with their associated element. } 
        \label{fig:Models-all}
\end{figure*}

Figure \ref{fig:Models-all} shows key lines in the WD SDSS observed spectrum (gray), as labeled, alongside the best fit model (red line) in 4 different regions.

\clearpage
\section{Linelist for Element Abundance Measurements in HD69962}
\label{app:Lines}
\begin{table}[ht]
\centering 
\begin{tabular}{llll}
\hline
Element & Wavelength (nm) & $\log(gf)$ & EP (eV) \\
\hline
Ca   1  & 610.2723 & -0.85  & 1.879 \\
Ca   1  & 612.2217 & -0.38  & 1.886 \\
Ca   1  & 643.9075 & 0.39   & 2.526 \\
Ca   1  & 644.9808 & -0.502 & 2.521 \\
Ca   1  & 647.1662 & -0.686 & 2.526 \\
Cr   1  & 527.5747 & -0.023 & 2.889 \\
Cr   1  & 529.6691 & -1.36  & 0.983 \\
Cr   1  & 534.8314 & -1.21  & 1.004 \\
Mg   1  & 552.8405 & -0.498 & 4.346 \\
Na   1  & 568.8205 & -0.404 & 2.104 \\
Na   1  & 615.4225 & -1.547 & 2.102 \\
Na   1  & 616.0747 & -1.246 & 2.104 \\
Ni   1  & 504.2186 & -0.58  & 3.658 \\
Ni   1  & 509.9930 & -0.10  & 3.679 \\
Ni   1  & 542.4645 & -2.77  & 1.951 \\
Ni   1  & 546.2493 & -0.818 & 3.847 \\
Ni   1  & 561.4773 & -0.573 & 4.154 \\
Ni   1  & 564.1881 & -1.046 & 4.105 \\
Ni   1  & 580.5217 & -0.579 & 4.167 \\
Ni   1  & 585.7752 & -0.402 & 4.167 \\
Ni   1  & 600.7310 & -3.40  & 1.676 \\
Si   1  & 570.1104 & -1.953 & 4.930 \\
Si   1  & 570.8400 & -1.370 & 4.954 \\
Si   1  & 577.2146 & -1.653 & 5.082 \\
Si   1  & 612.5021 & -1.464 & 5.614 \\
Si   2  & 634.7109 & 0.169  & 8.121 \\
Ti   2  & 501.3686 & -2.14  & 1.582 \\
Ti   2  & 506.9090 & -1.62  & 3.124 \\
Ti   2  & 507.2286 & -1.02  & 3.124 \\
Ti   2  & 518.8687 & -1.05  & 1.582 \\
Ti   2  & 522.6538 & -1.26  & 1.566 \\
Ti   2  & 541.8768 & -2.13  & 1.582 \\
\hline
\end{tabular}
 \caption{Elemental line data including ionization state, wavelength, $\log(gf)$ values, and excitation potential (EP).\label{tab:lines}}
\end{table}

\end{appendix}

\end{document}